\documentstyle[12pt,epsfig]{article}

\renewcommand{\theequation}{\arabic{section}.\arabic{equation}}

\def\hybrid{\topmargin -20pt    \oddsidemargin 0pt
        \headheight 0pt \headsep 0pt
        \textwidth 6.25in       
        \textheight 9.5in       
        \marginparwidth .875in
        \parskip 5pt plus 1pt   \jot = 1.5ex}
\hybrid
\def\be{\begin{equation}}       \def\eq{\begin{equation}}
\def\ee{\end{equation}}         \def\eqe{\end{equation}}

\def\bea{\begin{eqnarray}}      \def\eqa{\begin{eqnarray}}
\def\ena{\end{eqnarray}}        \def\eea{\end{eqnarray}}
                                \def\eqae{\end{eqnarray}}

\def\ba{\begin{array}}
\def\ea{\end{array}}
\def\unit{1 \hskip-.3em \raise2pt\hbox{$ \scriptstyle |$ } }

\def\a{\alpha}
\def\b{\beta}
\def\c{\gamma} 
\def\d{\delta}
\def\e{\epsilon}           
\def\f{\phi}               
\def\g{\gamma}
\def\h{\eta}   

\def\j{\psi}
\def\k{\kappa}                    
\def\l{\lambda}
\def\m{\mu}

\def\q{\theta}                    
\def\r{\rho}                                     
\def\s{\sigma}                                   
\def\t{\tau}

\def\D{\Delta}
\def\F{\Phi}
\def\G{\Gamma}

\def\L{\Lambda}
  
\def\P{\Pi}
\def\Q{\Theta}

\def\del{\partial}

\def\bop#1{\setbox0=\hbox{$#1M$}\mkern1.5mu
        \vbox{\hrule height0pt depth.04\ht0
        \hbox{\vrule width.04\ht0 height.9\ht0 \kern.9\ht0
        \vrule width.04\ht0}\hrule height.04\ht0}\mkern1.5mu}
\def\Box{{\mathpalette\bop{}}}                        
\def\pa{\partial}                              
\def\ddg{\sp\ddagger} 

\def\>{\rangle} 

\def\<{\langle} 
\def\Dsl{D \hskip-.6em \raise1pt\hbox{$ / $ } }

\def\leftrightarrowfill{$\mathsurround=0pt \mathord\leftarrow \mkern-6mu
       \cleaders\hbox{$\mkern-2mu \mathord- \mkern-2mu$}\hfill
       \mkern-6mu \mathord\rightarrow$}
\def\dvec#1{\vbox{\ialign{##\crcr
       \leftrightarrowfill\crcr\noalign{\kern-1pt\nointerlineskip}
       $\hfil\displaystyle{#1}\hfil$\crcr}}}          
\def\hook#1{{\vrule height#1pt width0.4pt depth0pt}}
\def\leftrighthookfill#1{$\mathsurround=0pt \mathord\hook#1
       \hrulefill\mathord\hook#1$}
\def\underhook#1{\vtop{\ialign{##\crcr                 
       $\hfil\displaystyle{#1}\hfil$\crcr
       \noalign{\kern-1pt\nointerlineskip\vskip2pt}
       \leftrighthookfill5\crcr}}}
\def\smallunderhook#1{\vtop{\ialign{##\crcr      
       $\hfil\scriptstyle{#1}\hfil$\crcr
       \noalign{\kern-1pt\nointerlineskip\vskip2pt}
       \leftrighthookfill3\crcr}}}

\def\sfrac#1#2{{\vphantom1\smash{\lower.5ex\hbox{\small$#1$}}\over
       \vphantom1\smash{\raise.4ex\hbox{\small$#2$}}}} 
\def\bfrac#1#2{{\vphantom1\smash{\lower.5ex\hbox{$#1$}}\over
       \vphantom1\smash{\raise.3ex\hbox{$#2$}}}}      
\def\afrac#1#2{{\vphantom1\smash{\lower.5ex\hbox{$#1$}}\over#2}}  
\def\on#1#2{{\buildrel{\mkern2.5mu#1\mkern-2.5mu}\over{#2}}}
\def\ddt#1{\on{\hbox{\LARGE .\kern-2pt.}}#1}             
\def\tdt#1{\on{\hbox{\LARGE .\kern-2pt.\kern-2pt.}}#1}   

\def\boxes#1{
       \newcount\num
       \num=1
       \newdimen\downsy
       \downsy=-1.5ex
       \mskip-2.8mu
       \bo
       \loop
       \ifnum\num<#1
       \llap{\raise\num\downsy\hbox{$\bo$}}
       \advance\num by1
       \repeat}
\def\boxup#1#2{\newcount\numup
       \numup=#1
       \advance\numup by-1
       \newdimen\upsy
       \upsy=.75ex
       \mskip2.8mu
       \raise\numup\upsy\hbox{$#2$}}

\newskip\humongous \humongous=0pt plus 1000pt minus 1000pt

\newif\ifdtup

\def\PRD{Phys. Rev. D}

\def\PRL#1#2#3{{\it Phys. Rev. Lett.} {\bf#1} (#2) #3}

\def\NPB#1#2#3{{\it Nucl. Phys.} {\bf B#1} (#2) #3}

\def\PRD#1#2#3{{\it Phys. Rev.} {\bf D#1} (#2) #3}

\def\PLB#1#2#3{{\it Phys. Lett.} {\bf #1B} (#2) #3}

\def\MPLA#1#2#3{{\it Mod. Phys. Lett.} {\bf A#1} (#2) #3}

\def\dda{\dot{\alpha}} 
\def\ddb{\dot{\beta}}
\def\ddc{\dot{\gamma}}
\def\ddd{\dot{\delta}}
\def\dde{\dot{\epsilon}}

\def\ddg{\dot{\gamma}}

\def\ddl{\dot{\l}}
\def\ddm{\dot{\m}}

\def\ddr{\dot{\r}}

\def\ddt{\dot{\t}}

\def\ta{\tilde{\a}}

\def\td{\tilde{\d}}

\def\ome{\omega}

\def\pa{\partial}  \def\pab{\bar{\pa}}
\def\del{\nabla}
\def\delbar{\bar{\nabla}}

\def\Hb{\bar{H}}

\def\Pb{\bar{\P}}
\def\ab{\bar{a}}

\def\ab{\bar{a}}
\def\bb{\bar{b}}

  \def\Tb{\bar{T}} 
   
  \def\Jb{\bar{J}}

\def\kbb{\bar{\kappa}} 

\def\fbb{\bar{\f}}

\def\ua{\underline{a}}
\def\ub{\underline{b}}
\def\uc{\underline{c}}
\def\ud{\underline{d}}
\def\ue{\underline{e}}
\def\uf{\underline{f}}
\def\ug{\underline{g}}

\def\uua{\underline{\a}}
\def\uub{\underline{\b}}
\def\uuc{\underline{\c}}
\def\uud{\underline{\d}}

\def\dif{\partial}
\def\difb{\bar{\partial}}
\def\dbar{\bar{\partial}}
\def\nonu{\nonumber \\{}}

\def\uug{\underline{\g}}

\begin{document}

\thispagestyle{empty}
\begin{flushright}
KUL-TF-96/26, LBNL-039753, UCB-PTH-96/59\\
hep-th/9704040
\end{flushright}
\vspace{1cm}
\setcounter{footnote}{0}
\begin{center}
{\LARGE{Self-Dual Supergravity from $N=2$ Strings}}\\[14mm]

{\sc Jan de Boer\footnote{e-mail:{\sc deboer@theor3.lbl.gov}}}\\[5mm]
{\it Department of Physics, University of California at Berkeley\\
         366 LeConte Hall, Berkeley, CA 94720-7300, USA\\
         and\\
         Theoretical Physics Group, Mail Stop 50A-5101\\
         Ernest Orlando Lawrence Berkeley National Laboratory, Berkeley, 
CA 94720, USA}\\[5mm]
{\sc and}\\[5mm]
{\sc Kostas 
Skenderis\footnote{e-mail:{\sc kostas.skenderis@fys.kuleuven.ac.be}}}\\[5mm]
{\it Instituut voor Theoretische Fysica,
         Katholieke Universiteit Leuven \\
         Celestijnenlaan 200D, B-3001 Leuven,
         Belgium}\\[20mm]

{\sc Abstract}\\[2mm]
\end{center}
A new heterotic $N=2$ string with manifest target space supersymmetry 
is constructed by combining a conventional $N=2$ string in the right-moving
sector and a Green-Schwarz-Berkovits type string in the left-moving
sector. The corresponding sigma model is
then obtained by turning on background fields 
for the massless excitations. We compute the beta functions and we
partially check the OPE's of the superconformal algebra 
perturbatively in $\a'$, all in 
superspace. The resulting field equations describe $N=1$ self-dual 
supergravity. 

\vfill

\newpage

\section{Introduction}
\setcounter{equation}{0}

Supergravity and string theories are intimately related to each other.
Supergravity theories, on the one hand, describe the
worldsheet theories
of strings and $p$-branes theories and, on the other hand, arise as 
the low energy theories of various superstring theories. Viewed 
differently, superstring theories serve as the definition 
of supergravity theories at the quantum level, as the field theories
describing supergravity theories themselves often
do not make sense at the quantum level.
There are supergravity theories, however, 
whose quantum description in terms of a string 
(or other) theory are not known. 
The most notable example
is eleven dimensional supergravity. The quantum theory having
eleven-dimensional supergravity as its low-energy effective field
theory has been 
coined $M$-theory. Recently, a description
of $M$-theory has been proposed \cite{matrix} in terms of the large $N$ limit
of a quantum-mechanical matrix model. 
A second outstanding example in this list of theories is four-dimensional
self-dual supergravity (SD SUGRA) \cite{kng1}. The search for a string theory 
that yields SD SUGRA as its low energy limit is the subject of this article.

Self-dual supergravity is interesting in its own right.          
Together with supersymmetric self-dual gauge theories
(SD SYM), it is believed to generate, upon dimensional reduction, 
all integrable models in two dimensions\cite{sdconj,gani}.
Furthermore, the role played by self-duality in four dimensions
seems to be similar to the role played by conformal symmetry 
in two dimensions \cite{moore}.

Self-dual theories in four dimensions are known to be closely related to
$N=2$ strings theories.  
These, though qualitatively quite different from $N=0$ and $N=1$ 
strings, have proven to constitute a rich family of string theories on 
their own, for reviews see e.g. \cite{k3,n1,bl2}.  After the introduction 
of the $N=2$ string in \cite{many1,many2,many3}
it took quite a while before it was shown in \cite{ov4,ov3,ov2} that the target
space theories described by $N=2$ strings are versions of self-dual gravity, 
possibly coupled to self-dual Yang-Mills. 

Recently, it was demonstrated in \cite{m4,m4,m2,m1} that
heterotic $(1,2)$ strings are also capable of having all 
known string theories as a background, so that it is in principle 
possible to recover arbitrary string theories by second quantizing 
$N=2$ strings. A remarkable fact about the
$N=2$ strings is that their spectrum consist of only a finite number 
of massless particles, a fact that may make the problem of developing a
string field theory tractable after all \footnote{An interesting discussion
of the string field theory description of the $N=2$ string can be
found in \cite{bersie2}.}. 
In addition, $(1, 2)$ strings seem to provide the low-energy 
world-volume description of $M$-theory. We shall argue in this article that
a new $N=2$ string, closely related to the $(2, 1)$ string, 
actually yields as a low energy effective theory $N=1$ self-dual 
supergravity. This leads to the conclusion that the two outstanding examples of
supergravity theories without known quantum definitions mentioned above
are actually related to each other.  
    
The conventional description of $(2, 1)$ strings has its drawbacks. 
Supersymmetry is not manifest and it is therefore quite hard to obtain the 
fermionic terms in the world-volume action of $M$ theory.
One has to resort to more 
cumbersome S-matrix techniques to get those terms. 
This led us to search for a more covariant formulation of 
$N=2$ strings. The basic idea is to start with an heterotic 
$N=(1,2)$ string theory, and to replace the $N=1$ sector 
by a Green-Schwarz-Berkovits (GSB) type sigma model. 
For four-dimensional compactifications of the heterotic
$N=(1,0)$ string, a similar procedure leads to a full super-Poincar\'e 
invariant sigma model, which enables one to derive the low-energy effective
action in superspace, keeping supersymmetry manifest all the time
\cite{bersie,ogp, ogpproc}. For $N=(1,1)$ strings, one can simultaneously
replace the left and right moving sectors by a Green-Schwarz-Berkovits 
type sigma model, yielding a covariant formulation of type II strings
\cite{bersie}. Here we extend this sequence to include heterotic 
$N=(1,2)$ strings.  

The Green-Schwarz-Berkovits (GSB) sigma model contains in addition
to the coordinates of 4d superspace an extra set of fields that
are the momenta conjugate to the anti-commuting coordinates
of 4d superspace. Different versions of $N=2$ strings that include 
these momenta in the world-sheet description have 
previously been considered in \cite{sez1, lp2}.
The idea here, however, is different than the one employed in \cite{sez1, lp2}.
Instead of replacing the conventional $N=2$ string 
with a new string theory, we combine an $N=2$ string in the right-moving
sector with a manifestly supersymmetric left-moving sector.
In other words, in our approach the right sector yields the 
self-duality and the left sector the supersymmetry.  
In addition, the sigma model includes a time-like chiral boson
$\rho$. At the same time, the $N=1$ algebra on the world-sheet
is enlarged to an $N=2$ algebra. This $N=2$ formulation is
related to the usual $N=1$ formulation via a non-local field
redefinition. Since each sector contains a differently realized $N=2$
world-sheet supersymmetry it seems appropriate to call this model
"type II heterotic". Because supersymmetry is kept manifest, there is
no need to perform a GSO projection. The heterotic string with 
$N=(1,2)$ world-sheet supersymmetry
has a left-moving world-sheet algebra which must
contain in addition to the $N=1$ super Virasoro algebra a
null current. In our case, we have to do the same, because
generally speaking the number of bosonic currents in the world-sheet
chiral algebra should be greater than or equal to the number of
timelike coordinates in target space. For the left-moving sector
this is three, two from the target space of signature $(2,2)$, and the
third one is the timelike boson $\rho$. We will choose a
particular null current involving the extra momenta. It is not
clear whether with this choice of null current the $N=2$ string
is equivalent to a known $N=2$ string, and whether other consistent choices
of null currents exist in our formulation.  

$N=2$ strings naturally live in a target space of dimension four 
with signature $(2,2)$. To write down the world-sheet action, 
one has to choose a complex structure in target-space, thereby 
breaking the Lorentz group from $SO(2,2)$ to $U(1,1)$. The group $U(1,1)$ 
is not able to  distinguish between different spins of particles, 
so that every physical state transforms in the same way as a scalar. 
Furthermore, $N=2$ strings have a $U(1)$ symmetry that make spectral flow 
into a gauge symmetry, making it difficult to determine even the 
target-space statistics of a physical state. For example, in \cite{s1}
it is claimed that $N=2$ strings lead to self-dual $N=4$ Yang-Mills and 
self-dual $N=8$ supergravity, whereas in \cite{ov4,ov3,ov2} only bosonic 
states were found (although the $Z_n$ models of \cite{ov3} are somewhat 
reminiscent of  theories involving superfields.) Also, in \cite{bkl1} 
fermions were found in the spectrum of $N=2$ strings, but they seem to 
decouple from the theory. To some extent, these problems may be related 
to the precise definition of the $N=2$ string one is employing, 
but for a better understanding of these issues a more covariant 
formulation of the $N=2$ string is clearly desirable. In \cite{bersie2}
it is suggested that in order to arrive at a covariant $N=2$ string field
theory, one has to attach Lorentz indices to the string field, but it is not
clear if this can be used to arrive at a more covariant world-sheet 
formulation.

A remarkable property of our formulation is that one can actually 
keep manifest $SO(2,2)$ Lorentz invariance in target space.  
This is possible because the GSB sector has manifest $SO(2,2)$ invariance in
target space.
By developing a sigma model that keeps these symmetries manifest we achieve
results that are manifestly $SO(2,2)$ covariant. This means, of course,
that in the process we sacrifice manifest worldsheet supersymmetry
which we now have to check by hand. 
We will check the closure of the GSB $N=2$ superconformal algebra
in the presence of curved background at tree and partially at
one-loop level 
and the closure of the conventional $N=2$ superconformal algebra
at tree level. Checking the superconformal algebra in the
GSB sector obviously does not break the target space Lorentz group.
It is not obvious, though, that this will be true for the calculations 
in the right sector as well. Indeed, even to write down the superconformal
generators one needs to choose a complex structure. So, any equation that 
follows from this sector will, in general, break the Lorentz invariance.
It turns out, however, that this sector supplies no equations at all,
and the $SO(2,2)$ Lorentz group remains unbroken.

The outline of this paper is as follows. In section two we describe the
new $N=2$ string propagating in a flat background, consider 
its spectrum and the linearized equations of motion for the target
space fields, and discuss their relation with self-dual supergravity. 
In these calculations the Lorentz symmetry is explicitely broken
to $U(1,1)$ simply by writing down the vertex operators.
Only the $\s$-model calculations are explicitly $SO(2,2)$ covariant. 
These are done in section three where we describe the new $N=2$ string
in a curved background, and confirm that the 4d sector of the
string does indeed yield self-dual supergravity. 
Finally in section four we discuss possible implications of our
results. The appendices describe our conventions and some more
technical details and results.



\section{A New $N=2$ String}
\setcounter{equation}{0}

\subsection{Flat Space}

The usual critical $N=(2,2)$ string propagating in a flat four-dimensional 
background
with signature $(2,2)$ is in superconformal gauge described by the following
action (conventions are given in appendix~A)
\be \label{act1}
S=\frac{1}{\a'} \int d^2 z
(\h_{a \ab} \pa \f^a \pab \f^{\ab} 
- {1 \over 2} \h_{a \ab} \pa \j^a \j^{\ab}
- {1 \over 2} \h_{a \ab} \bar{\pa} \chi^a \chi^{\ab}),
\ee
where $\phi^a$ is a complex boson, and $\psi^a,\chi^a$ are complex
fermions.
The heterotic $N=(1,2)$ string can be described by the same action plus
an action describing only left-moving degrees of freedom which we
denote by $S_{int}$. The left moving chiral algebra consists of an $N=1$
super Virasoro algebra and a super null current, with components with
conformal weight $1/2$ and $1$ which we denote by ${\cal J}_{null}^{1/2,1}$.
They are given by
\bea \label{nullc}
{\cal J}_{null}^{1/2} & = &  \lambda_a \chi^a + \lambda_{\ab} \chi^{\ab} 
+ {\cal J}^{1/2}_{int} 
\nonumber \\{}
{\cal J}_{null}^{1} & = &  i\lambda_a \pa \f^a 
+ i\lambda_{\ab} \dbar \f^{\ab} + {\cal J}^{1}_{int} 
\eea
and are required to have vanishing OPE's. There are two possibilities,
one can either take ${\cal J}_{int}^{1/2,1}$ to be zero and $\lambda^a$ to be 
a null
vector in $(2,2)$ dimensions, or one can take ${\cal J}_{int}^{1/2,1}$ 
non-zero.
Taking $\lambda^a=0$ does not lead to a unitary theory. The central charge 
$c_{int}$
of the internal theory $S_{int}$ should be chosen such that it cancels the 
central
charge $c=+6$ of the $4d$ matter fields and the central charges of the ghosts
for the left-moving chiral algebra, $c=-26+11-2-1$, yielding $c_{int}=12$.

Following \cite{ber1,ber2,ber3,bersie,ogp} we now replace the left-moving 
four-dimensional sector by a Green-Schwarz-Berkovits type world-sheet
theory. 
To make the right-moving $N=2$ supersymmetry manifest, we write
the action in $(0,2)$ superspace \cite{brmu}
with anti-commuting coordinates 
$\k,\kbb$.
Define the following $(0, 2)$ superfields
\bea
&&\F_a = \f_a + \k \psi_a + \k \kbb \difb \f_a \nonu
&&\F_{\ab}= \f_{\ab} + \k \psi_{\ab} - \k \kbb \difb \f_{\ab} \nonu   
&&\D_{\a} = p_{\a} + \kbb v_{\a} - \k \kbb \difb p_\a \nonu
&&\Q^\a = \q^\a + \k w^\a + \k \kbb \difb \q^\a \nonu
&&\D_{\dda} = p_{\dda} + \kbb v_{\dda} + \k \kbb \difb p_{\dda} \nonu
&&\Q^{\dda} = \q^{\dda} + \k w^{\dda} - \k \kbb \difb \q^{\dda}.
\eea 
We no longer see the left-moving fermions $\chi$, but have instead the
anti-commuting coordinates $\theta^{\alpha},\theta^{\dda}$ of 
4d superspace, their conjugate momenta $p_{\a},p_{\dda}$ and
some auxiliary fields. 
Out of these superfields we construct the following action
\be
S={1 \over 2\a'}\int d^2 z d^2 \k (\h_{a \ab} \F^a \difb
 \F^{\ab} - \D_\a \Q^\a 
+ \D_{\dda} \Q^{\dda}) - \frac{1}{2} \int d^2 z \pa \rho \dbar \rho
+S_{int},
\ee
where $\rho$ is a left-moving chiral boson that is inert under the right-moving
$N=2$ supersymmetry. In components this action takes the form
\bea  \label{act2}
S & = & \frac{1}{\a'} \int  d^2 z (\h_{a \ab} \pa \f^a \pab \f^{\ab} 
- {1 \over 2} \h_{a \ab} \pa \j^a \j^{\ab}
+ p_\a \pab \q^\a  + p_{\dda} \pab \q^{\dda}) \nonu & & 
- \frac{1}{2} \int d^2 z \pa \rho \dbar \rho
+{1 \over 2\a'}\int d^2 z (v_\a w^\a + v_{\dda} w^{\dda})
+S_{int}.
\eea
The commuting spinor fields $v_{\a}, v_{\dda}, w^\a, w^{\dda}$ 
are auxiliary and can be trivially integrated out. 
The left moving sector now coincides with that of the
Green-Schwarz-Berkovits sigma model, except for $S_{int}$. The 
four-dimensional part of the left-moving sector has an $N=2$ algebra,
with central charge $-3$, consisting of $+4$ from $\phi$, $-8$ from
$(d,\theta)$ and $+1$ from $\rho$. The $N=2$ ghosts contribute 
$-26+22-2=-6$. For unitarity, we still have to gauge a
super null current, and
the central charge of the internal sector is 
therefore given by $c_{int}=9-c_{null}$, where $c_{null}$ is the
central charge of the ghosts for the super null current.
The null currents we will consider have 
components of weight $1/2$ and $1$.
Such null currents can only exist if they form a
chiral or anti-chiral multiplet
with respect to the $N=2$ algebra.
Depending on whether the null current is bosonic or fermionic,
we need to
demand that $S_{int}$ has a $c=12$ or $c=6$ $N=2$ algebra
respectively. This is a stronger
condition than in the original heterotic $N=(1,2)$ string, where it
was only required to have a $c=12$ $N=1$ algebra. As the sigma model
above does have target space supersymmetry, the requirement that
the internal sector has an $N=2$ supersymmetry is probably equivalent
to the requirement of target space supersymmetry.
 The structure of the null current and internal sector,
and possible generalizations will be discussed below.  

Up to an overall factor of $1/\a'$, the right-moving $N=2$ algebra 
of (\ref{act2}) has generators
\bea
\Tb &=& - {1 \over 2} \h_{\ua \ub} (\pab \f^{\ua} \pab \f^{\ub}
+\j^{\ua} \pab \j^{\ub}) \\
H &=& i \sqrt{2} \h_{a \bb} \j^a \pab \f^{\bb} \\
\Hb &=& i \sqrt{2} \h_{\ab b} \j^{\ab} \pab \f^{b} \\
\Jb &=& \h_{\ab b} \j^{\ab} \j^{b}
\eea
whereas the left-moving $N=2$ algebra has generators
\bea 
J & = & -i\dif\rho + J_{int} \nonu
G & = & \frac{1}{i\a' \sqrt{8\a'}} e^{i\rho} d^{\a} d_{\a}  + G_{int} \nonu
\bar{G} & = &  \frac{1}{i\a' \sqrt{8\a'}} e^{-i\rho} d^{\dda} d_{\dda} 
+\bar{G}_{int}
\nonu
T & = & 
 \frac{1}{\alpha'} (
 -\frac{1}{2} \h_{\ua \ub} \pa \f^{\ua} \pa \f^{\ub}
  - p_{\a} \dif \theta^{\a}
- {p}_{\dda} \dif \theta^{\dda} )
+\frac{1}{2} \dif \rho \dif \rho + T_{int}.
\eea
Here, we defined
\bea d_{\alpha} &  = &  p_{\alpha} + i\theta^{\dda} \dif \phi_{\a\dda} + 
 \frac{1}{2} \theta^{\dda} \theta_{\dda} \dif \theta_{\a} - 
 \frac{1}{4} \theta_{\a} \dif( \theta^{\dda} \theta_{\dda}) \nonu
d_{\dda} &  = &  p_{\dda} + i\theta^{\a} \dif \phi_{\a\dda} + 
 \frac{1}{2} \theta^{\a} \theta_{\a} \dif \theta_{\dda} - 
 \frac{1}{4} \theta_{\dda} \dif( \theta^{\a} \theta_{\a}).
\eea
The target space supersymmetry generators are given by
\bea 
q_{\a} & = & \oint \frac{dz}{2\pi i} \left( p_{\a} 
- i\theta^{\dda} \dif \phi_{\a\dda} +
  \frac{1}{4} \theta_{\a} \dif (\theta^{\dda} \theta_{\dda})  \right)  \nonu
q_{\dda} & = & \oint \frac{dz}{2\pi i} \left( p_{\dda} 
- i\theta^{\a} \dif \phi_{\a\dda} +
  \frac{1}{4} \theta_{\dda} \dif (\theta^{\a} \theta_{\a})  \right).
\eea
Notice that these generators only act on the left-moving sector,
and they (trivially) commute with the right moving sector.
In particular, both $\bar{\pa} \f^{\a \dda}$ and $\psi^{\a \dda}$ 
are (trivially) 
inert under the target space SUSY. 
One can check that $d_\a$ and $d_{\dda}$ are also inert. 
To make the target space SUSY manifest we further define 
the following variables that also commute with target space SUSY 
\bea
\P_{\a \dda} &=& \pa \f_{\a \dda} - i \q_\a \pa \q_{\dda} + 
i \pa \q_\a \q_{\dda} \nonu 
\P_{\underline{\a}} &=& \pa \q_{\underline{\a}}
\eea
These new variables are in fact given by
\be
\P^A = \pa z^M E_M{}^A
\ee
where $z=\{\f^{\m \ddm}, \q^\m, \q^{\ddm}\}$ are the superspace coordinates
and $E_M{}^A$ are the vielbeins of flat superspace
that can be used to convert curved 
indices into flat indices. The vielbeins are one  along the diagonal 
and the only non-zero off-diagonal elements are
\be \label{eq02}
E_{\ddm}{}^{\a\dda} = i \delta_{\ddm}{}^{\dda} \theta^{\a},
\qquad E_{\m}{}^{\a\dda} = i \delta_{\m}{}^{\a} \bar{\theta}^{\dda} .
\ee

With these variables one may rewrite the flat space model in a manifestly
supersymmetric form,
\be \label{act3}
S=\frac{1}{\alpha '} \int d^2 z (
\frac{1}{2} \Pi^{\a\dda} \bar{\Pi}_{\a\dda}
+ d_{\a} \bar{\Pi}^{\a}
+ d_{\dda} \bar{\Pi}^{\dda}
+\frac{1}{2} \bar{\Pi}^A \Pi^B B_{BA}
-\frac{\a'}{2} \difb \rho \dif \rho
- {1 \over 2} \h_{a \ab} \pa \j^a \j^{\ab}) + S_{int}
\ee
where we introduced an anti-symmetric tensor field $B_{BA}$ whose
only non-zero components are
\be \label{eq03}
\begin{array}{cc} B_{\a\dda,\ddb} = i C_{\ddb\dda} \theta_{\a} &
                  B_{\a\dda,\b} = i C_{\b\a} \bar{\theta}_{\dda} \\
                  B_{\ddb,\a\dda} = -i C_{\ddb\dda} \theta_{\a} &
                  B_{\b,\a\dda} = -i C_{\b\a} \bar{\theta}_{\dda} 
\end{array}
\ee

Covariant derivatives are in flat superspace given by
$\nabla_A=E_A{}^M \dif_M$, where the inverse vielbein $E_A{}^M$ has
one on the diagonal and off-diagonal components
\be E_{\dda}{}^{\m\ddm} = -i \delta_{\dda}{}^{\ddm} \theta^{\m},
\qquad E_{\a}{}^{\m\ddm} = -i \delta_{\a}{}^{\m} \bar{\theta}^{\ddm},
\ee
as follows from (\ref{eq02}). From this we deduce that the 
non-zero torsion (see appendix A)
of flat superspace is in our conventions given by
\be T_{\a,\dda}{}^{\b\ddb}=-2i \delta_{\a}{}^{\b} \delta_{\dda}{}^{\ddb}.
\label{flattor}
\ee
Associated to the anti-symmetric tensor field in (\ref{eq03}) is
a non-zero field strength $H_{ABC}$. With curved indices it
is given by $H_{MNP}=\frac{1}{4} \partial_{[M} B_{NP)}$ which
in terms of flat indices becomes
\be H_{ABC} = \frac{1}{4}  \nabla_{[A} B_{BC)} -\frac{1}{4}
T_{[AB}{}^D B_{|D|C)}.
\ee
The only non-zero components of $H$ are
\be
H_{\a,\dda,\b\ddb}=-i C_{\a\b} C_{\dda\ddb}
\label{flath}
\ee
and its permutations.

It remains to define the null current for (\ref{act2}) to make the 
theory complete.
It seems very complicated to keep track of the null current in the field 
redefinition
that leads one from the RNS to the Green-Schwarz-Berkovits formulation. 
Therefore,
our approach will be to find a null current that satisfies the required 
properties. As 
mentioned above, the null currents should be part of a chiral or 
anti-chiral multiplet
of the $N=2$ algebra. In addition, they should preserve target space 
supersymmetry.
This more or less uniquely
 fixes the null currents to be of the form
\bea \label{nullcurr}
{\cal J}_{null}^{1/2} & = &  e^{i\rho} u^\a d_\a \nonu
{\cal J}_{null}^{1} & = & u^\a \Pi_\a.  
\eea
The spinor $u^{\alpha}$ can be either bosonic or fermionic. If it is
bosonic, the central charge of the internal sector has to be
$c_{int}=6$, if it is fermionic $c_{int}=12$. Naively, one would
think that the internal sector has to be a meromorphic conformal
field theory, in order to have a modular invariant partition function.
No such theories with $c=6$ are known;  for $c=12$ one
particular example has been constructed in \cite{pierce}, 
see also \cite{m4,m3,m2,m1}. It consists of 8 fermions and 
8 bosons compactified on an $E_8$ lattice. Therefore, one would
be more inclined to consider only the case where $u^{\alpha}$ is
fermionic. On the other hand, from the point of view of the generators
(\ref{nullcurr}) a bosonic $u^{\alpha}$ seems more natural. In addition,
the modular transformation properties of the four-dimensional part
of the theory are already quite intricate, and it is not clear
what the precise conditions on the internal sector are in order
to guarantee modular invariance. We will therefore in the remainder
consider the case of a bosonic spinor $u^{\alpha}$, and discuss the
modifications that have to be made for a fermionic spinor in section~4.

A bosonic spinor $u^{\alpha}$ is automatically null, and gives rise to
the required 
vanishing OPE between ${\cal J}_{null}^{1/2}$ and ${\cal J}_{null}^{1}$.
The latter contains a term $\sim e^{i\rho} u^\a u_\a /(z-w)^2$.
Such terms cannot be cancelled by anything from the 
internal sector. Therefore, (\ref{nullcurr}) cannot be modified
so as to include any contribution from the internal sector. 
It is at present not clear whether there is a, perhaps different,
possibility to construct null currents that do include
contributions from the internal sector. Alternatively,
one could imagine introducing a null current that forms
an unconstrained representation of the world-sheet $N=2$
algebra, in which case the internal 
sector would have to be modified. An example of a full null current
that does not break target space SUSY is given by  
$U^{\a \dda} (\P_{\a \dda}, d_\a \P_{\dda} e^{i \r}, d_{\dda} \P_\a e^{-i \r}, 
\P_\a \P_{\dda})$, where $U^{\a \dda}$ is a null vector.
This is the supersymmetrization of the usual 
choice of null current in conventional $N=2$ strings, our chiral
null current being sort of its square root. It turns out, however,
that introducing such a full null current trivializes the theory. 


\subsection{Vertex Operators}

We now turn to a description of the spectrum of the $N=2$ string.
The spectrum of $N=2$ strings has been studied in detail, see
e.g. \cite{ov4,ov3,ov2,lp1,bkl1,l4,l3}. 
These results need to be combined with the results for the spectrum
of the Green-Schwarz-Berkovits \cite{ber3,bersie}.
The spectrum of the pure $N=2$
string consists of one scalar, the Lee-Yang scalar that
describes self-dual gravity. In our case this scalar field
is promoted to a scalar superfield $W(x,\theta)$. Due to the
presence of the null-current, it is not an unconstrained superfield.
The constraints will be discussed below. 
Let $G,\bar{G}$ be the 
world-sheet supersymmetry generators of the left-movers,
and $H,\bar{H}$ the world-sheet supersymmetry generators of the
right-movers. Then the 4-D integrated vertex 
operators corresponding to $W(x,\theta)$ reads
\be \label{VO1}
V=\int d^2 z \{G,\{\bar{G},\{H^+,\{H^-,W(x,\theta) \}\}\}\},
\ee
where $\{A, B\}$ denotes the single pole in the OPE of $A$ with $B$.
In addition, there are states in the theory coming from the internal
sector. These are described by integrated vertex operators of the form
\be \label{VO2}
V^{i}=\int d^2 z \{ H^+, \{H^-, \{G,M^{i}_1 \Omega\} 
+ \{\bar{G},M^{i}_2 \bar{\Omega} 
 \}\}\}
\ee
where $M^{i}_1(x,\theta)$ is a real chiral superfield, $M^{i}_2(x, \theta)$ 
is a real anti-chiral superfield (in space with signature (2,2) one can have
real chiral superfields \cite{kng1}) and
$\Omega$, $\bar{\Omega}$ are chiral (anti-chiral) primaries of the internal
$N=2$ superconformal algebra with conformal weight $1/2$. 
If the internal sector is represented by two free $N=2$ superfields,
there are 2 chiral 
and 2 anti-chiral primaries of conformal weight 1/2, and $i$ runs from 1 to 2.
For other internal sectors, needed e.g. for alternative choices of 
null current, the index $i$ may take on a different set of values. 
We  will suppress the superscript $i$ for the time being.
The $W,M_1,M_2$ are regular superfields in target-space,
and are not allowed to depend on $\psi^a$.
The on-shell conditions for the vertex operators yield the 
following linearized equations of motion,
\be 
D^2 W = \bar{D}^2 W = \Box W = 0 \label{eq4D}
\ee
and 
\be
D^2 M_1 = \bar{D}^2 M_2 = 0. \label{internal}
\ee
Here, $D_\a, D_{\dda}$ are the standard (rigid) covariant derivatives
in superspace. 
In addition, the vertex operators (\ref{VO1}) and (\ref{VO2}) have
certain gauge invariances. The ones associated with the right-moving 
sector imply that $W$, $M_1$ and $M_2$ can be chosen not to 
depend on $\bar{\P}^A$, a fact already used above, 
whereas the ones associated with the left-moving sector read
\be
\delta W = \Lambda + \bar{\Lambda}
\ee
where $\L$ and $\bar{\L}$ are chiral and anti-chiral superfields respectively. 

This is not yet the full story, because the vertex operators should also
preserve the world-sheet
gauge invariance generated by the null currents. The spin-one
component of the null current generates the transformation 
$\delta d^{\a} = \epsilon u^{\a}$ and leaves all other world-sheet
fields invariant. Thus $\delta \{\bar{G},\{G,W(x,\theta)\}, \}\sim
\{\bar{G}, e^{i\rho} u^{\a} D_{\a} W(x,\theta) \}$, and this
vanishes if 
\be
u^{\a} D_{\a} W(x,\theta)=0.
\label{nul}
\ee
In a similar way we find that $M_1$ should satisfy
\be 
u^{\a} D_{\a} M_1(x,\theta)=0,
\label{null}
\ee
and that there is no new equation emerging for $M_2$.
In addition, the Hilbert space of the theory should be identified with its
spectral flow generated by the $U(1)$ current. This leads to extra 
gauge invariances\footnote{The null vector is associated 
with a first class constraint
and, therefore, should "kill twice", once by imposing the constraint and once
by the gauge invariance it generates} 
\be \delta W(x,\theta) = u^{\a} D_{\a} Y(x,\theta), 
\qquad \delta M_1(x,\theta) = u^{\a} D_{\a} Y_1(x, \theta) 
\label{nuga} \ee
with $Y_1$ such that it preserves the chirality of $M_1$. 

Let us first analyze the internal sector. The second equation 
in (\ref{internal})
imply that the component fields of real anti-chiral superfield
$M_2$ satisfy the usual linearized field equations.
We shall now show that the equation 
(\ref{null}) together with the gauge invariance in (\ref{nuga}) completely
eliminate $M_1$.
Indeed, equation (\ref{null}) is solved by
\be
M_1=u^\a D_\a N
\ee
where $N$ is such that $M_1$ is chiral but otherwise arbitrary.
One may clearly choose $Y_1$ such that $M_1$ is gauged away.

Let us now consider the 4D vertex operator. One can go from the gauge variant 
prepotential $W$ to a gauge invariant field strength $W_{\uua}$ in the usual way
\be
W_\a = \bar{D}^2 D_\a W; \qquad W_{\dda} = D^2 \bar{D}_{\dda} W
\ee
Equations (\ref{eq4D}) imply
\be
D^2 W_\a =0; \ \ \bar{D}^2 W_{\dda} =0, \ \ \Box W_{\uua}=0.
\label{equa4D}
\ee
In addition, we have the equation (\ref{nul})
that now becomes
\be u^\a W_\a = 0 \label{null2} \ee
This equation is solved by
\be
W_\a = u_\a X
\ee
where $X$ satisfies the equations that follow from (\ref{equa4D}) but it is 
otherwise arbitrary.
The null gauge invariance (\ref{nuga}), in turn, reads
\be
\d W_\a = {1 \over 2} u_\a \bar{D}^2 D^2 Y; \ \ 
\d W_{\dda} = -2i u^\a \pa_{\a \dda} D^2 W = 0
\label{nuga1}
\ee   
{}From this one can derive that the null gauge invariance can
be used to gauge away $W_\a$.
A somewhat more detailed analysis of both the spacetime and 
internal sector in components can be found in appendix B.

We, thus, see that in the $4D$ sector the linearized field equation 
are given by the equation $W_\a=0$. This equation describes $U(1)$ 
self-dual SYM theory \cite{kng1}, and as such contains 
$1+1$ degrees of freedom. The fermionic one is the Majorana-Weyl fermion
$\l_{\dda}$
and the bosonic one is the Yang-Lee scalar describing the self-dual
field strength. We shall shortly see that these degrees of freedom can 
be consistently
identified with one helicity component of the gravitino and graviton, 
respectively.

To summarize, the spectrum consist of 1+1 components coming from the 4D part,
and $q(1+1)$ real components from $M_2$, where $q$ is the number of
(anti)chiral primaries of weight $1/2$ of the internal sector.
There is no conventional 
target space action  that can describe these degrees of freedom since 
one cannot write down a fermionic kinetic term for any of these multiplets. 

Let us now show that the equations of motion we have found are consistent
with those of self-dual supergravity. For this, 
we want to show that the $N=1$ SD SUGRA equation $W_{\a \b \g} = 0$
(but $W_{\dda \ddb \ddg} \neq 0$) \cite{kng1}
reduces after we choose a complex structure, at the linearized level and
in a specific gauge and Lorentz frame to the equation 
$W_\a=0$ (but $W_{\dda} \neq 0$) satisfied by the vertex operators.

The gravitino field strength at the linearized level is given
by the following formula 
(up to irrelevant numerical factors; see \cite{book}, equation (5.2.4))
\bea
W_{\a \b \g}&=& \bar{D}^2 D_{(\a} \pa_\b{}^{\ddb} \hat{H}_{\g) \ddb}, \\
W_{\dda \ddb \ddg}&=& 
D^2 \bar{D}_{(\dda} \pa^\b{}_{\ddb} \hat{H}_{|\b| \ddg)}, 
\label{gravitino}
\eea
where $D_\a, \bar{D}_{\dda}$ are the flat superspace covariant 
derivatives (we are working at the linearized level) and $\hat{H}_{\ua}$ 
is the SUGRA prepotential (in \cite{book} this prepotential is denoted
by $H_{\ua}$).
These equations are invariant under the following gauge transformations
\be
\d \hat{H}_{\ua} = \pa_{\ua} \l
\ee 
We fix this gauge invariance by setting
\be
\pa^{\ua} \hat{H}_{\ua} =0 \label{fix}
\ee
One can solve this relation by expressing 
$\hat{H}_a$ is terms of a prepotential $B_{\ua \ub}$,
\be 
\hat{H}_{\ua} = \pa^{\ub} B_{\ub \ua},
\ee
Expressing $B_{\ua \ub}$ (which is antisymmetric in $\ua, \ub$) 
in terms of its Lorentz irreducible (i.e. symmetric in their indices)
components $B_{\a \b}$ and $B_{\dda \ddb}$ in the usual way we get
\be
\hat{H}_{\a \ddb} = \pa_{\a}{}^{\ddg} B_{\ddb \ddg} + \pa^\g{}_{\ddb} B_{\a \g}
\label{prepre}
\ee
Then, the gravitino field strength takes the form
\bea
W_{\a \b \g}&=& 
\bar{D}^2 D_{(\a} \pa_\b{}^{\ddb} \pa_{\g)}{}^{\ddg} B_{\ddb \ddg}
- {1 \over 2} \Box \bar{D}^2 D_{(\a} B_{\b \g)}, \\
W_{\dda \ddb \ddg}&=& 
D^2 \bar{D}_{(\dda} \pa^\b{}_{\ddb} \pa^{\g}{}_{\ddg)} B_{\b \g}
- {1 \over 2} \Box D^2 \bar{D}_{(\dda} B_{\ddb \ddg)}.
\eea

Let us now consider the specific Lorentz frame in which only
$p^1-p^3 \neq 0$ ($p^1$ and $p^3$ are real). 
Such a choice is possible since we are on-shell.
In complex coordinates (with the complex structure of appendix A), 
we have
\be
p_{+\dot{+}}=-p_{-\dot{-}}=-i p_{+\dot{-}}=-i p_{-\dot{+}}
\ee 
One may now work out $W_{\a \b \g}$ and  $W_{\dda \ddb \ddg}$. The 
result is 
\bea 
W_{+++} &=& -3 \bar{D}^2 D_{+} W \nonu
W_{++-} &=& -\bar{D}^2 D_{-} W - 2i \bar{D}^2 D_{+} W \nonu
W_{+--} &=& \bar{D}^2 D_{+} W - 2i \bar{D}^2 D_{-} W \nonu
W_{+++} &=& 3 \bar{D}^2 D_{-} W 
\eea
where 
\be
W= 2p_{+\dot{+}}^2 (B_{\dot{+}\dot{+}}-B_{\dot{-}\dot{-}}
+2iB_{\dot{+}\dot{-}})
\ee
To get the result for $W_{\dda \ddb \ddg}$ one simply replaces dotted
by undotted indices and vice versa. 
One may verify that the reality of $\hat{H}_{\ua}$ implies 
\be
\overline{B_{++}}=-B_{--}; \ \  \overline{B_{+-}}=-B_{+-}
\ee
and similar results for $B_{\dda \ddb}$.
With these reality properties $W$ is real.

Let us now define 
\be
W_\a = \bar{D}^2 D_\a W
\ee
(one can similarly define $W_{\dda}$).
Obviously the condition $W_{\a \b \g}=0$ implies precisely 
$W_\a=0$ and in addition $W_{\dda \ddb \ddg}$ and $W_{\dda}$ are not
set equal to zero. Hence, the SD equation $W_{\a \b \g}=0$ 
after choosing a complex structure, at the linearized 
level and in a specific gauge and Lorentz frame, is precisely the equation
satisfied by the VO's.

Finally, we give the relation between the components
of the SUGRA multiplet and the components of the linearized analysis
(the components of the SUGRA are given in \cite{book}, equation (5.2.8))
\bea
&&\l_{\dda} = D^2 \bar{D}_{\dda} W \big| 
= 2 p^{\a}{}_{\dot{-}} D^2 \bar{D}_{\dda} \hat{H}_{\a \dot{-}} \big| \sim 
p^{\a}{}_{\dot{-}} \psi_{\ua, \dot{-}} \\
&&f_{\a \b} = D_{(\a} W_{\b)} \big| =
4i p_{-}{}^{\dda} p_{(\a}{}^{\ddb} 
[D_{\b)}, D_{\ddb}] \hat{H}_{- \dda} \big|
\sim p_{-}{}^{\dda} p_{(\a}{}^{\ddb} h_{\b)\ddb, -\dda} \\
&& D=D^\a W_\a \big| = 
2 p_{-}{}^{\ddb} D^\a \bar{D}^2 D_{\a} \hat{H}_{- \ddb} \big|
\sim p_{-}{}^{\dda} A_{- \dda}
\eea
where $h_{\a \dda}$ is the conformal graviton, $\psi_{\a \dda, \b}$ is the 
conformal gravitino and $A_{\ua}$ is the auxiliary vector of conformal 
supergravity. (Notice that the projection definition of $A_{\ua}$ 
involves also a term proportional to $\e_{\ua \ub \uc \ud}$. 
This term, however, drops out in our gauge).

\section{The New $N=2$ String in an Arbitrary Background}
\setcounter{equation}{0}

So far we have focused on the vertex operators of the theory, and
the linearized equations of motion they satisfy. In this section
we will couple the new $N=2$ string to an arbitrary curved
background. In order for the background to be compatible with the
world-sheet symmetries, several constraints have to be satisfied.
The main result will be that the constraints are exactly those
that describe self-dual supergravity, confirming the results from
the previous section. 

The action (\ref{act3}) can easily be put in a curved background, 
by simply taking the vielbeins and anti-symmetric tensor 
to be arbitrary rather than those of flat superspace. Upon expanding them
to first order around a flat background we should recover the massless
vertex operators of the previous section. However, one then sees that
(\ref{act3}) is not yet complete,
because the massless vertex operators do contain
terms bilinear in the right-moving fermions $\psi$, and we
need to include such terms in the action as well. 
These terms are completely fixed by requiring the right-moving
sector to have $N=2$ supersymmetry, and we find for the action in a curved
background (with the background fields
coming from the internal sector turned off)
\bea \label{act4}
S & =&  \frac{1}{\a'} \int d^2 z ( \frac{1}{2} 
\eta_{\ua \ub} \Pi^{\ua} \bar{\Pi}^{\ub} + 
 d_{\a} \bar{\Pi}^{\a} + d_{\dda} \bar{\Pi}^{\dda} +
\frac{1}{2} \bar{\Pi}^A {\Pi}^B B_{BA} \nonu
&  & \qquad -\frac{\a'}{2} \dbar \rho
 \dif \rho + \eta_{\ua \ub} \psi^{\ua} \nabla \psi^{\ub} 
+  \Pi^{\ua} \j^{\ub} \j^{\uc} T_{\uc \ub \ua} + d^{\uua} \j^{\ub} \j^{\uc} 
T_{\uc \ub \uua}) 
+ S_{int}.
\eea
Here, $\nabla$ is the pull-back of the target space covariant
derivative to the world-sheet,
\be
\del \psi^{\ua} = \dif \psi^{\ua} + \P^B [\ome_{B \b}{}^{\g} M_{\g}{}^{\b}
+ \ome_{B \ddb}{}^{\ddg} M_{\ddg}{}^{\ddb} + \G_B Y,\psi^{\ua}],
\ee
where $\omega$ and $\Gamma$ are the spin and $U(1)$ connections in
target space. 
One could also imagine adding a term 
$\P^{\uua} \j^{\ub} \j^{\uc} T_{\uc \ub \uua}$ to the action. 
However, such a term is incompatible with the local target space 
$U(1)$ invariance given by
\be
\delta \Pi^{\ua} = 0,
\delta \Pi^{\a} = - \frac{1}{2} \Lambda \Pi^{\a}, 
\delta \Pi^{\dda} =  \frac{1}{2} \Lambda \Pi^{\dda},
\delta d_{\a} = \frac{1}{2} \Lambda d_{\a},
\delta d_{\dda} = -\frac{1}{2} \Lambda d_{\dda},
\delta \psi^{\ua}=0
\label{targu1}
\ee
and, in addition, breaks $N=2$ world-sheet supersymmetry.

The generators of the world-sheet chiral and anti-chiral
algebras read as follows. The generators of the left-moving $N=2$
are
\bea 
J & = & -i \dif \rho \nonu
G & = &  \frac{1}{i \alpha' \sqrt{8\alpha'}} e^{i\rho}
 d^\a d_\a 
 \nonu
\bar{G} & = &  \frac{1}{i \alpha' \sqrt{8\alpha'}} e^{-i\rho}
  d^{\dda} d_{\dda} 
  \nonu
T & \rightarrow & 
 \frac{1}{\alpha'} \left(
 -\frac{1}{2} \Pi^{\a\dda} \Pi_{\a\dda}  - d_{\a}  \Pi^{\a}
- {d}_{\dda}  \Pi^{\dda} 
+ \frac{\alpha'}{2} \dif \rho
   \dif \rho
\right). 
\label{e-gen}
\eea
The generators of the null current are still of the form
(\ref{nullcurr}),
\bea 
{\cal J}_{null}^{1/2} & = &  e^{i\rho} u^\a d_\a \nonu
{\cal J}_{null}^{1} & = & u^\a \Pi_\a.  
\eea
The generators of the right-moving $N=2$ algebra read  
\bea
\Tb &=& - {1 \over \a'}( \eta_{\ua \ub} (\frac{1}{2} \bar{\Pi}^{\ua} \bar{\Pi}^{\ub}
+   \j^{\ua} \bar{\nabla} \j^{\ub})   +\bar{\Pi}^{\ua}\j^{\ub} \j^{\uc} T_{\uc\ub\ua})   \nonu
H^+ &=& \frac{i}{\sqrt{2} \a'} \left(
(\h_{\ua \ub} + I_{\ua \ub}) \j^{\ua} \bar{\Pi}^{\ub}  
+ \j^{\ua} \j^{\ub} \j^{\uc} C^+_{\uc \ub \ua} \right) \nonu
H^- &=& \frac{i}{\sqrt{2} \a'} \left(
(\h_{\ua \ub} - I_{\ua \ub}) \j^{\ua} \bar{\Pi}^{\ub}  
+ \j^{\ua} \j^{\ub} \j^{\uc} C^-_{\uc \ub \ua} \right) \nonu
\Jb &=& \frac{1}{2 \a'} I_{\ua \ub} \j^{\ua} \j^{\ub}.
\label{fgen}
\eea
where $I_{\ua \ub}$ is the complex structure that satisfies
\be
I_{\ua \ub} = - I_{\ub \ua}; \qquad I_{\ua \ub} I^{\ub \uc} = \d_{\ua}{}^{\uc}
\label{cpxconv}
\ee
and $C^+_{\ua \ub \uc}$, $C^-_{\ua \ub \uc}$ are tensors to be determined
and that vanish in the flat space limit. One may easily check that in the 
flat space limit this algebra correctly reduces to the correct
$N=2$ algebra.

We will first analyze the constraints coming from $N=2$ world-sheet
supersymmetry, postponing the analysis of the null current to section~3.3.
We find that the left-moving $N=2$ supersymmetry imposes exactly
the same set of constraints as in the case of the heterotic string 
(see \cite{ogp}). The right movers yield only one constraint, namely
$\del_{\uug} I_{\ua\ub}=0$, and fix $C^{\pm}_{\ua \ub \uc}$ to be of 
the form (\ref{aux330}). This is a highly non-trivial calculation,
the details of which follow below. 

The equations of motion of  (\ref{act4}) read
\bea
0&=& \bar{\P}_{\uua}- \j^{\ub} \j^{\uc} T_{\uc \ub \uua}=0  \label{deq}
\label{eqmot1} \\
0&=&\nabla \j_{\ua} + \Pi^{\ub} \j^{\uc} T_{\uc \ua \ub} - 
d^{\uud} \j^{\uc} T_{\uc \ua \uud}
\label{eqmot2}
\eea
The field equation for the fields $\P^{\ua}$ and $d^{\uua}$ differ from the
field equations we found in \cite{ogp} by terms bilinear in the 
fermion fields $\j^{\ua}$. This means that the tree level constraints
that follow from the left-moving sector are the ones found in \cite{ogp}
plus possibly some new constraints due to the extra terms. The former 
ones supplemented by a maximal set of conventional constraints 
was solved in \cite{ogp} to yield the following supergravity algebra
\bea
\{\del_{\a}, \del_{\b}\} &=& 0,
\label{bian1} \\
\{\del_{\a}, \del_{\ddb} \}
&=& -2i \del_{\a \ddb} -4i H_{\ddb \g} M_{\a}{}^{\g}
+4i H_{\ddg \a} M_{\ddb}{}^{\ddg} +4i H_{\ddb \a} Y, 
\label{bian2} \\
\left[\del_{\a}, \del_{b} \right] &=&
-2 \del_{\b} H_{\ddb \g} M_{\a}{}^{\g}
\nonumber \\
&\ &+[-2i C_{\a \b} W_{\ddb \ddg}{}^{\ddd}
+ C_{\ddb \ddg} (\del_{(\a} H^{\ddd}{}_{\b)} -
\frac{1}{3} C_{\a \b} \del^{\e} H^{\ddd}{}_{\e})] M_{\ddd}{}^{\ddg}
\nonumber \\
&\ &+ 2 \del_{\b} H_{\ddb \a} Y  \label{bian3} \\
\left[\del_{a}, \del_{b} \right] &=&
\{-2 H_{\dda \b}  \del_{\a \ddb} \nonumber \\
&\ &+[\frac{i}{2} C_{\a \b} \del_{(\dda} H_{\ddb)}{}^{\g}
+C_{\dda \ddb}(-\frac{i}{6} C^{\g}{}_{(\a|} \del^{\dde} H_{\dde| \b)}
+ W_{\a \b}{}^{\g})] \del_{\g} \nonumber \\
&\ &+\left[ C_{\dda \ddb} \left(\frac{1}{24} \del_{(\a} W_{\b \g}{}^{\d)}
+\frac{1}{4}
(C^{\d}{}_{\a} \del_{(\b|\dde} H^{\dde}{}_{|\g)} + \a \leftrightarrow \b)
\right. \right.
\nonumber \\ 
&\ &
\left. \left.
+\frac{i}{6} C_{\a \g} C^{\d}{}_{\b} \del^{\dde} \del^{\e} H_{\dde \e}
\right)
+\frac{i}{2} C_{\a \b} \del_{\g} \del_{(\dda} H_{\ddb)}{}^{\d}\right]
M_{\d}{}^{\g} \nonumber \\
&\ &-\frac{i}{2} C_{\a \b} \del^{\d} \del_{(\dda} H_{\ddb)\d} Y + {\rm c.c.} \}
\label{bian4}
\eea
where `c.c.' denotes our definition of complex conjugation (not to be confused
with the usual complex conjugation which in a space of signature
$(2,2)$ does not mix dotted and undotted indices), 
which acts as follows on the various objects
\be
(\del_{\a})^{\dagger}  =  \del_{\dda} ;\quad \quad
(C_{\a\b})^{\dagger}  =  C_{\dda\ddb} ;\quad \quad
(M_{\gamma}{}^{\delta})^{\dagger}  =  M_{\ddg}{}^{\ddd}  ;\quad \quad
(Y)^{\dagger}  =  -Y  \label{conjug1} ;
\ee\be
(\del_{\ua})^{\dagger}  =  -\del_{\ua}  ;\quad \quad
(H_{\ddb\a} )^{\dagger}  =  H_{\dda\b}  ;\quad \quad
(W_{\a\b\g})^{\dagger}  =  W_{\dda\ddb\ddg}. 
\label{conjug2}
\ee
and, in addition, does not reverse the order of factors, namely 
$(AB)^{\dagger}=A^{\dagger} B^{\dagger}$.
$W_{\a \b \g}$ is a completely symmetric chiral superfield,
\be
\del_{\ddd} W_{\a \b \g} = 0,
\ee
and $H_{\dda \b}$ is
defined in terms of $H_{ABC}$ by 
\be
H_{a b c} = C_{\g \a} C_{\ddg \ddb} H_{\dda \b} -
C_{\g \b} C_{\ddg \dda} H_{\ddb \a}. \label{habc}
\ee
$W_{\a \b \g}$ and $H_{\dda \b}$ satisfy the following differential
relations
\be
\del_{\ua} H^{\ua} = 0, \
\del^{\g} W_{\g \a \b} = \frac{i}{6} \del_{(\a|} \del^{\ddg} H_{\ddg| \b)}
+\frac{i}{2} \del^{\ddg} \del_{(\a|}  H_{\ddg| \b)}, \
\del^{\b} \del_{\b} H_{\ua} = 0.
\ee
Furthermore, all the components of the field strength $H_{ABC}$ vanish
except the ones that are given in (\ref{habc}) and (\ref{flath}).
Compactly, we have
\be
T_{ABc} + (-1)^{AB} 2 H_{ABc} = 0.
\ee 

The remaining field equations now take the form
\bea 
\bar{\del} d_{\a} &=&
\j^{\ub} \j^{\uc} 
[\P^{\ud} (R_{\ud \a \ub \uc} + 2i T_{\uc \ub \ddd} C_{\a \d}
+ 2 \del_{\a} H_{\uc \ub \ud}) \nonu
&& \qquad + \P^{\ddd} (R_{\ddd \a \ub \uc} -4i H_{\uc \ub, \a \ddd}) 
+ d^{\uud} \del_{\a} T_{\uc \ub \uud}] \nonu
\del \bar{\P}_{\ua} &=& -2 \P^{\uc} \Pb^{\ub} H_{\ub \uc \ua}
- \Pb^{\uc} T_{\uc \ua}{}^{\uub} d_{\uub}  \nonu
&&+\j^{\ub} \j^{\uc}
[\P^{\ud} (R_{\ua \ud \ub \uc} + 2 H_{\ue [\ua \ub} H_{\uc]\ud \ue}
+2 \del_{\ud} H_{\uc \ub \ua} - 2 \del_{\ua} H_{\uc \ub \ud})
\nonu
&& \qquad +\P^{\uud} (R_{\ua \uud \ub \uc} + 2 \del_{\uud} H_{\uc \ub \ua})
\nonu
&&\qquad +d^{\uud} (\del_{\ua} T_{\uc \ub \uud} 
- 4 T_{\ub \ue \uud} H_{\uc \ue \ua})]
\label{eqmot3}
\eea

We want to determine whether (\ref{act4}) is indeed invariant under
the symmetries generated by (\ref{e-gen}), (\ref{fgen}).
First, we check whether the terms bilinear
in fermions bring in any new equations in the analysis of the
left sector. 
Using the $d$ field equation and ignoring the terms which are proportional 
to  $d^2$ as these terms can be removed by
modifying the transformation rule of the gravitini 
that couple to the supersymmetry currents (see
\cite{ogp}) we get that the following three equations should hold
\bea
&&R_{\ud \a \ub \uc} + 2i T_{\uc \ub \ddd} C_{\a \d}
+ 2 \del_{\a} H_{\uc \ub \ud} =0  \label{Dcond1} \\
&&R_{\ddd \a \ub \uc} -4i H_{\uc \ub, \a \ddd} = 0 \label{Dcond2} \\
&&\del_{\a} T_{\uc \ub \ddc} = 0. \label{Dcond3}
\eea
A direct computation, which consist of substituting the relevant expressions
{}from the supergravity algebra for the tensors involved  
shows that all of them are automatically satisfied. This means that 
the extra terms bilinear in the fermions produce no new equations.

We now move to the analysis of the right moving sector.
Conservation of the $U(1)$ current $\pa \bar{J}=0$ implies the
following three equations
\bea
&&\del_{\uuc} I_{\ua \ub} = 0 \label{Jcond1} \\
&&\del_{\uc} I_{\ua \ub} + 2 I_{\ud [\ua} H_{\ub] \uc \ud}=0 \label{Jcond2} \\
&&I_{\ud [\ua} T_{\ub] \ud \uuc} = 0 \label{Jcond3}
\eea
The last two conditions actually follow from the first one as we 
now show. Equation (\ref{Jcond2}) simply follows from the 
equation 
\be
\{ \del_{\c}, \del_{\ddc} \} I_{\ua \ub} = 0.
\ee
Differentiating (\ref{Jcond2}) by $\del^{\ddc}$ yields 
equation (\ref{Jcond3}).  
To get this result one may use the following identities 
\bea
&&[\del_{\dda}, \del_{\ub}] V_\a = -2i T_{\ua \ub}{}^\g V_\g, \label{id1} \\
&&[\del_{\a}, \del_{\ub}] V_\g = 2 C_{\g \a} \del_\b H_{\ddb}{}^\d V_\d, 
\label{id2}\\
&&T_{\a\dda, \b\ddb, \g} - T_{\g\dda, \b \ddb, \a} = 
i C_{\a \g} \del_{\ddb} H_{\dda \b} \label{id3}
\eea
In these identities $V_{\uua}$ is assumed to have the $U(1)$ charge of its 
index, namely $1/2$ if the index is undotted and $-1/2$ if the index is 
dotted. One may use the first two identities to derive how 
commutators act on tensors.
(One has to be careful, however, when the $U(1)$ charge of the tensor is 
different from the one its indices indicate.) 

The condition in (\ref{Jcond2}) is the familiar condition that the complex 
structure should be covariantly constant with respect to a connection
with torsion. In our case, however, this is a derived condition!
As is usual in supersymmetric theories, there is a more fundamental
spinorial condition in lower (mass) dimension. This condition 
states that the complex structure is both chiral and anti-chiral.
In flat space this condition correctly implies that the complex structure
is constant. 

Having analyzed the $U(1)$ current we now turn to the two supersymmetry
currents. The equation $\pa H^{\pm}=0$ involves terms linear in the 
fermions $\j^{\ua}$ and terms cubic in the fermions $\j^{\ua}$.
The ones linear in the fermions vanish upon using the equations
(\ref{Jcond2}) and (\ref{Jcond3}). The ones cubic in the fermion 
field yield six equations that contain two unknowns, the 
tensors $C^{\pm}_{\uc \ub \ua}$. It is an excellent consistency check that 
this system of equations does have a unique solution. 
The computations involved are similar to the ones we already presented so far 
but considerably more complicated. The details can be found in appendix C. 
Here we only give the final solution for $C^{\pm}_{\uc \ub \ua}$,
\be \label{aux330}
C^{\pm}_{\uc \ub \ua} = \frac{1}{3}(-2 H_{\uc \ub \ua} \pm 
I_{\ud [\uc} H_{\ub \ua] \ud}) 
\ee

This almost concludes
the discussion of the anti-holomorphic $N=2$ algebra. We
only have to show that the stress-energy tensor $\bar{T}$ in
(\ref{fgen}) is conserved, $\dif \bar{T}=0$. For this we do
not have to do any complicated analysis, we simply observe
$\bar{T}$ is the Noether current for the symmetry $\bar{z}
\rightarrow \bar{z} + \bar{\epsilon}(\bar{z})$, which automatically
guarantees $\dif\bar{T}=0$. 

\subsection{The dilaton}

In the previous section we have discussed how to couple the $N=2$ string to an
arbitrary curved background, except for the dilaton. Although we have not 
worked out the precise form of the 
dilaton vertex operator, it can be included in the action
using a suitable generalization of the Fradkin-Tseytlin term \cite{bersie,ogp}.
The form of this term can be determined by looking at the geometry of a
super world-sheet. In the case of the heterotic string one has to consider a
super world-sheet with $N=(2,0)$ supersymmetries, in the case of  the heterotic
$N=2$ string we have a holomorphic and anti-holomorphic $N=2$ algebra and the
relevant world-sheet geometry is that of $N=(2,2)$ superspace. 
In the formulation
of 
\cite{n22gra} 
there are four types of  world-sheet supercurvatures,
that are (anti)-chiral and twisted (anti)-chiral from the world-sheet point 
of view.
In order to write down a supersymmetric world-sheet action, these four types
of curvatures have to couple to target space fields with the same world-sheet
properties. (Anti)-chirality for the holomorphic $N=2$ algebra translates 
directly
into (anti)-chirality for the target space fields, as one sees from the OPE's
$\{G,\phi\}\sim \nabla_{\a} \phi, \{\bar{G},\phi\}\sim \delbar_{\dda} \phi$.
Similarly, the OPE's show that (anti)-chirality with respect to 
the anti-holomorphic $N=2$ translates into the property $(\eta_{\ua \ub} 
\pm I_{\ua\ub}) \del_{\ub} \phi=0$. These are the (anti)-holomorphic
derivatives in target space, and we see that a world-sheet chiral superfield
is a target space holomorphic chiral superfield, a world-sheet twisted chiral
superfield is a target space anti-holomorphic chiral superfield, etc. 
We denote the sum of the target space holomorphic and anti-holomorphic
chiral superfields by $\phi$, and the sum ot the target space holomorphic
and anti-holomorphic anti-chiral superfields by $\bar{\phi}$. 
Once the form of the Fradkin-Tseytlin term has been determined, one can go
to superconformal gauge. Then there is no dilaton-dependent term in the
action anymore, except for a coupling between the $\rho$-field and the dilaton,
 and the only place the dilaton appears is in the generators
of the $N=2$ algebras. The dilaton contributions to the holomorphic $N=2$
algebra are identical to the ones derived in \cite{ogp}. The contributions to 
the anti-holomorphic $N=2$ generators can be derived in a similar way, but
since we will not need those results we will not present them here.

After inclusion of the dilaton, the action still has the local $U(1)$ 
invariance
given in (\ref{targu1}),  if one also varies $\rho,\phi,\bar{\phi}$ as follows
\be
\delta \phi=-\frac{1}{2} \Lambda, \qquad
\delta \bar{\phi}=\frac{1}{2} \Lambda, \qquad
\delta \rho= i \Lambda.
\ee

As in \cite{ogp}, it will be convenient to redefine $\rho$ so that it becomes 
a $U(1)$-invariant quantity, by defining
\be \label{rhoredef}
\rho \rightarrow \rho - i(\phi-\bar{\phi}).
\ee
After this redefinition, $\rho$ itself is a chiral boson that does not 
couple to
any of the other fields in the theory, and it can be quantized exactly. We
need to do so in order to perform an expansion in $\a'$ in the theory; the
field $\rho$ does not carry a factor of $\alpha'$ in the action, and therefore
an arbitrary number of loops for the $\rho$-field contributes to the 
theory at a given order in  $\alpha'$.

In the presence of an additional null current in the superconformal algebra
it is in principle possible to introduce another dilaton-like field, 
that couples
in a Fradkin-Tseytlin term to the curvature of the $U(1)$ gauge field needed
to gauge the null current. We have not examined whether or not it is consistent
to introduce such a field but it would be interesting to know if it is at all
possible, and whether there is a corresponding state in the cohomology of
the $N=2$ string.

\subsection{The internal sector}

Until now we have suppressed the internal sector. One may turn on 
the corresponding background fields by 
including terms in the sigma model of the form\cite{bersie}
\be
S_{int} = \int  d^2 z \{ H^+, \{ H^-, \{ \bar{G} ,M^{i} \bar{\Omega}
    \} \} \}
\label{intcou}
\ee  
where $M^{i}$ is set of $q$ real anti-chiral superfields and 
$H^{\pm}, \bar{G}$ are the generators of the left and right superconformal   
algebra in a curved background (\ref{e-gen})-(\ref{fgen}).
Including these terms into the action partially changes the analysis 
of the previous section since now the worldsheet field equations
receive contributions from the internal sector and there are additional
vertices to be taken into account into our 1-loop and beta function 
computation. One may easily verify that the new terms in the 
field equations contain a factor $\exp i\r$.
This factor is essential so that one does not run into problems
already at tree level. The same factor, however, presents the most 
serious obstacle in computing the contribution of the internal sector
to the one loop results. We have been using a hybrid 
method that treats the $\r$  field exactly and the rest of the fields
perturbatively in $\a'$. This seems quite difficult in the presence of the 
internal sector since now the $\r$ field does not decouple from the
rest of the fields. We shall not discuss the internal sector
in the remainder of this section.

\subsection{The null current}

So far our results were generic for any heterotic $(1,2)$ string, 
regardless of the
choice of null current. In this section
we analyze what further constraints we have to impose on the background in 
order to 
incorporate the null current (\ref{nullcurr}) into the theory. 
The null currents, after incorporating the dilaton and the redefinition of 
the $\rho$-field,
read
\be
{\cal J}_{null}^{1/2} = e^{i\rho} e^{\phi-\bar{\phi}} u^{\a} d_{\a},
\qquad
{\cal J}_{null}^{1} =  u^{\a} \Pi_{\a}.
\ee

First, we examine the conditions imposed by requiring that the null currents 
be holomorphic.
In fact, as in \cite{ogp}, we will only require the weaker condition that 
$\dbar {\cal J}$ is
proportional to ${\cal J}$. The weaker condition guarantees that, at tree 
level, we can still gauge the
$N=2$ algebra together with the null current, and the theory is well-defined. 
Using the
equations of motion for the background fields (\ref{eqmot1}), (\ref{eqmot2}), 
(\ref{eqmot3})
and the auxiliary identity
\be
\nabla \bar{\Pi}^A - \delbar \Pi^A = -\Pi^B \bar{\Pi}^C T_{CB}{}^A
\ee
we find the following set of equations
\bea
u^{\a} T_{\uc \ub \a} & = & 0 \nonu
\nabla_{\ub} u_{\a} & = & 0 \nonu
u^{\a} \nabla_{\d} T_{\uc\ub \a} - (\nabla^{\uug} u_{\d}) T_{\uc \ub \uug} 
& = & P_{\uc \ub} u_{\d} \nonu
T_{\uc \ub}{}^{\uud} \nabla_{\uud} u^{\a} 
  - u^{\d} \del_{\d} T_{\uc \ub}{}^{\a} 
& = & Q_{\uc \ub} u^{\a}
\label{cons1}
\eea
for some tensors $P_{\uc \ub}$, $Q_{\uc \ub}$. 
These equations are not all independent. 
The second equation in (\ref{cons1})
implies in particular $[\del_{\ub},\del_{\uc}]u_{\a}=0$, and this implies 
both the third equation
with $P_{\ub \uc}=0$, and the fourth equation, with 
$Q_{\uc\ub}\sim F_{\uc\ub}$. 

\subsection{Background Field Expansion}

To further analyze the theory we will compute some of the OPE's between the 
various generators.
The techniques are identical to those described in \cite{ogp}. The only new 
ingredient is
the presence of the fermions $\psi^{\ua}$. For these, the background field 
rules read
\bea
&&\D \psi^{\ua} = 0, \label{psibg}\\
&&\D (\del \psi^{\ua}) = \P^B y^C R_{C B}{}^{\ua \ud} \psi_{\ud} .
\eea
After performing the background field expansion, we will give the fermions
a background expectation value by replacing $\psi^{\ua} \rightarrow
\psi^{\ua} + \psi^{\ua}_{bg}$. In addition to the background field expansion
presented in \cite{ogp}, there are now two kinds of additional terms. One kind
comes from the terms containing fermions $\psi_{\ua}$, the other from terms 
containing $\bar{\Pi}^{\uua}$. 
The latter is no longer zero but rather satisfies the field
equation given in (\ref{eqmot1}). 
Loops are computed as in \cite{ogp} by treating the fermions
$\psi^{\ua}$ in the same way as $d^{\a},y_{\a}$.

Using the techniques described above, 
we examine the tree-level OPE between $G$ and ${\cal J}_{null}$. We will 
not consider
the contributions of the dilaton to the tree-level OPE. As discussed in 
\cite{ogp}, in many
cases these contributions are ambiguous and can be put equal to zero either 
by adding total
derivatives to the action, or by modifying the background field expansion 
of the $d$-field. 
The tree-level OPE between $G$ and ${\cal J}^1_{null}$ should have a single 
pole which
is proportional to $\partial({\cal J}^{1/2}_{null})$. A straightforward 
calculation shows that
this is only true if 
\be 
\nabla u_{\a} = - \nabla_{\a} u^{\b} \Pi_{\b}.
\ee
{}From this we deduce two new equations for $u_{\a}$, 
namely $\del_{\ddb} u_{\a}=0$
and $\del_{(\a} u_{\b)}=0$. Further constraints can be found by considering 
the OPE between 
${\cal J}^{1/2}_{null}$ and ${\cal J}^1_{null}$. That OPE contains a single 
pole with a
coefficient proportional to $u^{\a} (\del_{\a} u^{\b}) \Pi_{\b}$. This can 
only vanish if
$\del_{\a} u_{\b}=0$. Putting everything together, we have derived that 
$\del_A u_{\b}=0$,
i.e. $u_{\b}$ should be a covariantly constant bosonic null spinor. This is 
the most natural
generalization of the constant spinor $u_{\b}$ to curved superspace. By 
computing
$[\del_A,\del_B]u_{\g}=0$ we can now derive further consistency conditions 
that the background
fields have to satisfy. One finds
\be
\del_A u_{\b} = 0 \Longrightarrow H^{\dda\b} u_{\b} = 
T_{\ua\ub}{}^{\g} u_{\g} = 0.
\ee

{}From here on, one could in principle proceed and 
compute further OPE's in order to check the rest of the
$N=2$ algebra. Before doing so, it is useful to first impose as many 
constraints as possible
on the background fields in order to simplify the calculations. In the 
analysis of the
vertex operators we used spectral flow to argue that certain operators could
 be identified
with each other. In particular, if one obtains the constraint 
$u^{\a} \del_{\a} M=0$ from
the null current, the identification under spectral flow can be used to 
choose $\del_{\a} M=0$.
Spectral flow acts on vertex operators by multiplication by 
$\exp(\omega \int^z {\cal J}^1_{null})$
which in flat space reduces to $\exp(\omega u^{\a} \theta_{\a})$. It acts on 
each vertex
operator independently, and one can choose a different parameter $\omega$ for 
each choice
of momentum of the vertex operator.
These expressions clearly indicate the difference between the $RNS$ 
formulation of the (1,2) strings and our new formulation.
In the former the $U(1)$ current generates a gauge invariance that 
kills one (or two) spacetime bosonic coordinates. In our case
the null current effectively kills one fermionic coordinate instead.
The vertex operators contain prepotentials, whereas the
action contains potentials, and it is quite hard to see precisely how 
spectral flow
affects the various background fields, connections, etc. Our philosophy will 
be that if the
null-current gives rise to a constraint of the form 
$u^{\a} ({\rm something})=0$, then one should
be able to choose (${\rm something}=0$) as well. 
One advantage of this is that there will be
no constraints that explicitly depend on $u^{\a}$ anymore, and as in the case 
of the vertex
operators the null current imposes a covariant set of constraints. In 
particular, we will
choose the constraints
\be T_{\ua \ub}{}^{\gamma}= H_{\dda\b} = 0
\label{finalconstr} \ee
{}from now on.
Indeed, these equations are Lorentz covariant and do not depend on the 
choice of $u^\a$. This reflects the fact that Majorana-Weyl fermions
exist in $2+2$ dimensions and, therefore, one can effectively kill one
fermionic direction without breaking the Lorentz group. By imposing the
constraints (\ref{finalconstr}) the only nonzero fields left in the
theory are the dilaton and $W_{\dda \ddb \ddg}$. The supergravity algebra
obtained after imposing (\ref{finalconstr}) reads
\bea
\{\del_{\a}, \del_{\b}\} &=& 0, \nonu 
\{\del_{\a}, \del_{\ddb} \} &=& -2i \del_{\a \ddb}, \nonu
\left[\del_{\a}, \del_{\ub} \right] &=& 
-2i C_{\a \b} W_{\ddb \ddg}{}^{\ddd} M_{\ddd}{}^{\ddg}, \nonu
\left[\del_{\dda}, \del_{\ub} \right] &=& 0,\nonu 
\left[\del_{\ua}, \del_{\ub} \right] &=&
C_{\a \b}(W_{\dda \ddb}{}^{\ddg} \del_{\ddg} + 
{1 \over 24} \del_{(\dda} W_{\ddb \ddg}{}^{\ddd)} M_{\ddd}{}^{\ddg}),
\label{sdsugra}
\eea
So far, we ignored the condition that the complex structure has to be 
covariantly
constant, and one may worry that this condition imposes further constraints on
the background fields. The complex structure can be written as
\be
I_{ab} = C_{\a\b} \bar{I}_{\dda \ddb} + C_{\dda \ddb} I_{\a\b},
\ee 
where $I_{\a\b}$ and $I_{\dda\ddb}$ are symmetric. In order for $I$ to 
square to
the identity matrix, we need that either $I_{\a\b}=0$ or
$\bar{I}_{\dda\ddb}=0$. In our case, the most natural choice that is compatible
with (\ref{sdsugra}) is to take $\bar{I}=0$. One immediately verifies that 
covariant constancy of $I_{\a\b}$ does not lead to any further constraints
on the background fields, which validates our original claim that the 
right-moving
part of the new $N=2$ string does not lead to any new equations. 

One may wonder whether or not it is possible to relax some of the conditions 
given in
(\ref{finalconstr}) and still arrive at a consistent string theory. It would 
certainly
be interesting to investigate this further. Therefore, when some of the 
results presented below
are to our knowledge
still valid after relaxing some of the conditions in (\ref{finalconstr}), 
we will make a
comment to this effect. We already notice, however, that the
number of degrees of freedom left after imposing (\ref{finalconstr}) is 
in agreement with the analysis of the vertex operators.

\subsection{Further tree-level and one-loop results}

As a further consistency check of this $N=2$ string we have made a partial 
check
of the OPE's of the holomorphic $N=2$ algebra. After imposing 
(\ref{finalconstr}),
the tree-level OPE's between the null currents come out correct. Notice that 
due to
the presence of the $\rho$-field in ${\cal J}^{1/2}_{null}$, which satisfies 
the
OPE $e^{i\rho(z)} e^{i\rho(w)} \sim e^{2i\rho(w)}/(z-w) + \ldots$, we have to 
keep
terms with up to two background fields in its OPE (counting $\psi^{\ua}$ as 
$1/2$ background 
field). The tree-level diagrams for the OPE's of $G,\bar{G}$ with $G,\bar{G}$ 
and the null
current also yield the right results.
If we drop the condition $H_{\dda\b}=0$ 
but keep $T_{\ua\ub\g}=0$, the only
vertex in the action contributing to the OPE of 
$G$ with ${\cal J}^{1/2}_{null}$ 
itself is 
$-2i \int d^2 z D_{\a} y^{\b} y_{\b} \bar{\Pi}^{\ddb} H_{\ddb}{}^{\a}$.
The result is proportional to $H_{\ddb}{}^{\a}u_{\a}$, which vanishes. 
If we also
drop the constraint $\del_{\a} u_{\b}=0$ but keep $\del_{(\a} u_{\b)}=0$,
the OPE of $G$ with ${\cal J}^{1/2}_{null}$ can still work if we add a suitable
total derivative to the action and use the identity 
$\del_{\dda} \del^{\b} u_{\b}
\sim H_{\dda}{}^{\b} u_{\b}$. 

At one-loop level we also considered the 
diagrams that contribute to the OPE's of
$G,\bar{G},{\cal J}_{null}$ with themselves. Again, everything works out
 correctly.
If one assumes $T_{\ua\ub}{}^{\g}=0$ but keeps $H_{\dda\b}$ unconstrained, 
there are
still many simplifications. For example, the action contains no vertices
$\int d^2 (\del y^{\a}) y^B \j^{\uc}_{bg} \j^{\ud}_{bg}$ or 
$\int d^2 y^{\a} (\del y^B ) \j^{\uc}_{bg} \j^{\ud}_{bg}$, and the OPE's of
$G,\bar{G}$ with themselves are still OK. The other one-loop diagrams are more
complicated and it seems at first sight quite unlikely that there would be 
any chance that their sum vanishes unless we impose $H_{\dda\b}=0$.

We have not computed any further OPE's. In the case of the heterotic string 
we found
the field equations from the OPE of $T$ with $G$ and $T$ with $T$, and 
noticed that
these can at the same time be derived using a conventional beta-function 
calculation.
Here, we will assume the same relation holds, and proceed by doing a much 
faster
beta-function calculation. The results will show that the background has to 
be Ricci-flat.
This is the usual condition in order to have a conventional $N=2$ algebra, 
strongly
suggesting that the anti-holomorphic $N=2$ algebra will also persist at 
one-loop.
We will therefore not perform any diagrammatic analysis to check the 
anti-holomorphic
algebra either.

\subsection{Beta-function calculation}

The conventional beta-function calculation can be performed along the same 
lines
as explained in \cite{ogpproc}. The idea is to find all UV divergent 
contributions to
the effective action at one-loop, and to cancel the resulting conformal 
anomaly using
the dilaton. In contrast to the case of 
the heterotic string discussed in \cite{ogpproc}, there
are several different one-loop diagrams that contribute. The new diagrams 
all have one
background $\Pi^A$ or $D^{\uua}$ field sticking out, and two background 
$\psi^{\ua}$'s, but not
all at the same vertex.
All of them can be worked out in a straightforward fashion using dimensional 
regularization.
The only point that has to be treated with some care is the fact the some 
diagrams have
cancelling UV and IR divergences, and in order to isolate the UV divergence 
we have to
first subtract out the IR divergence. The results, not assuming any 
constraints coming
{}from a null current, are given in appendix~D. Here we will discuss the 
results specialized
to the case where we impose in addition $T_{\ua\ub\g}=H_{\dda\b}=0$. 

Inserting $T_{\ua\ub\g}=H_{\dda\b}=0$ into the results in appendix~D yields 
the equations
\bea
0 & = &  \frac{1}{2} \del^{\uc} R_{\uc\ua\uf\ug} 
  + \frac{1}{2} \del_{\ua} T_{\ug\uf}{}^{\dda} \del_{\dda} 
(\phi+\bar{\phi}) \nonu
  & & +\frac{1}{2} T_{\ug\uf}{}^{\ddb} \del_{\ua} \del_{\ddb} (\phi+\bar{\phi})
  + \frac{1}{2} R_{\ud\ua\uf\ug} \del_{\ud} (\phi+\bar{\phi}) 
\label{ab1} \\{}
0 & = &  \frac{1}{2} \del^{\uc} R_{\uc\a\uf\ug} 
  +\frac{1}{2} \del_{\ta} T_{\ug\uf}{}^{\ddb} \del_{\ddb} (\phi+\bar{\phi}) 
\nonu
  & & -\frac{1}{2} T_{\ug\uf}{}^{\ddb} \del_{\a} \del_{\ddb} (\phi+\bar{\phi})
  + \frac{1}{2} R_{\ud\a\uf\ug} \del_{\ud} (\phi+\bar{\phi})
\label{ab2} \\{}
0 & = & T_{\ug\uf}{}^{\ddb} \bar{\nabla}^2 (\phi+\bar{\phi})
 \label{ab3} \\{}
0 & = &   \frac{1}{2} \del^{\uc} \del_{\uc} T_{\ug\uf\dda} 
    + \frac{1}{2} R_{\uc\ub\ug\uf} T_{\uc\ub\dda} 
   -\frac{1}{2}
    T_{[\uf|\ub\dda|} T_{\ug]\ub}{}^{\ddg} \del_{\ddg} (\phi+\bar{\phi}) \nonu
& &   +\frac{1}{2} \del_{\ud} T_{\ug\uf\dda} \del_{\ud} (\phi+\bar{\phi})
  - T_{[\uf|\ue\dda|} H_{\ug]\ue\ud} 
\del_{\ud} (\phi+\bar{\phi}) 
\label{ab4} \\{} 
0 & = &\del_{\b} \del_{\ua} (\phi+\bar{\phi})
\label{ab5} \\{}
0 & = & \del_{\ddb} \del_{\ua} (\phi+\bar{\phi}).
\label{ab6}
\eea

Quite interestingly,
using (\ref{ab5}) and (\ref{ab6}) one can show that all the dilaton terms in
(\ref{ab1}), (\ref{ab2}) and (\ref{ab4}) vanish. For example, (\ref{ab5}) and
(\ref{ab6}) imply $\del_{\ub} \del_{\ua} (\phi+\bar{\phi})=0$ and by
antisymmetrizing in $\ua$ and $\ub$ we find $T_{\ua\ub}{}^{\ddg} \del_{\ddg}
(\phi+\bar{\phi})=0$. From here we conclude that $\del_{\ddg}
(\phi+\bar{\phi})=0$. Similarly, $[\del_{\uc},\del_{\b}]\del_{\ua} 
(\phi+\bar{\phi})=0$ implies $R_{\ud\a\ub\uc} \del_{\ud} 
(\phi+\bar{\phi})=0$, and $[\del_{\uc},\del_{\ub}]\del_{\ua} 
(\phi+\bar{\phi})=0$ implies $R_{\ud\ua\ub\uc} \del_{\ud} 
(\phi+\bar{\phi})=0$, etc. Equation (\ref{ab3}) is now satisfied since
the dilaton,  in addition to being chiral, also satisfies 
$\del^2 \f + \del^\a \f \del_\a \f =  
\bar{\del}^2 \fbb + \bar{\del}^{\dda} \fbb \bar{\del}_{\dda} \fbb =0$.
These
relations can be seen by examining the OPE of G with itself and also follow
{}from the low-energy effective action presented in \cite{ogp}.

The precise interpretation of these statements
is not entirely clear. We already noticed that $\phi+\bar{\phi}$ is a sum
of a target space holomorphic and anti-holomorphic function. The results
above show it is also the sum of a world-sheet holomorphic and 
anti-holomorphic
function. The fact that the dilaton decouples from the field equations may 
either
mean that it is always forced to be a constant, or that it plays the role of 
some
kind of Lagrange multiplier in the theory. It would be very interesting to see
if this can be related to the results of \cite{bl1} who showed that
in order to restore target space covariance in the $N=2$ string one has to
let the dilaton transform in a non-trivial way under target space Lorentz 
transformations.
We have not performed  a detailed analysis of Lorentz anomalies in our $N=2$
string, but it is possible that the one-loop anomalies can only be canceled by
assigning a non-trivial transformation rule to the dilaton. We leave this 
issue to
a future discussion and will for the time being simply drop the dilaton from
our considerations. 

Without the dilaton, the following short list of equations remains
\bea
0 & = &   \del^{\uc} R_{\uc\ua\uf\ug} 
\label{ac1} \\{}
0 & = &   \del^{\uc} R_{\uc\a\uf\ug} 
\label{ac2} \\{}
0 & = &   \del^{\uc} \del_{\uc} T_{\ug\uf\dda} 
+ R_{\uc\ub\ug\uf} T_{\uc\ub\dda}
\label{ac3}
\eea
We can work this further out be expressing everything in terms 
of $W_{\dda\ddb\ddg}$,
using (\ref{sdsugra}). Before doing so, notice that (\ref{sdsugra}) 
in particular implies that 
$R_{\ua\ub\uc\ud} \sim C_{\a\b} C_{\g\d} \del_{(\dda} W_{\ddb\ddg\ddd)}$, so
that the Ricci  tensor vanishes, as we expected from the presence 
of the ``conventional'' anti-holomorphic $N=2$ algebra. 

\begin{sloppypar}
Coming back to (\ref{ac1})-(\ref{ac3}), we find that (\ref{ac1}) is a direct
consequence of  $\{ \del^{\b}, \del^{\ddb} \} W_{\ddb\ddg\ddd}=0$,
and that (\ref{ac2})  follows from 
$[\del^{\b},\bar{\del}^2] W_{\ddg\ddd\dda}=0$.
Equation (\ref{ac3}) is the only one that yields a non-trivial result, namely
\be
-\frac{1}{4} \del^{\dda} (W^{\ddr}{}_{(\dda \ddb} W_{\ddg \ddd) \ddr} ) 
+ 2 \del_{\dda} (W_{\ddr \ddb \ddg} W^{\dda \ddr}{}_{\ddd}) = 0 .
\ee
This implies finally
\be \label{ad1}
\del^{\ddg} (W^{\ddr}{}_{\ddg(\ddd} W_{\ddl )\dda \ddr} )=0.
\ee
This equation is the only additional piece of information 
that we have in addition to the 
supergravity algebra with $W_{\a \b \g} = H_{\ua} =0$.
This equation may still receive corrections from the internal sector.
\end{sloppypar}

\section{Discussion}
\setcounter{equation}{0}

We have constructed in this article  
a "heterotic type II" $N=2$ string theory that has 
manifest target space supersymmetry,
and we have shown by computing the beta functions and by 
checking the OPE's of the superconformal algebra 
perturbatively that its low energy theory is $N=1$ self-dual supergravity.
We have, thus, solved the problem of finding a 
consistent quantum theory that has SD SUGRA as its low energy limit.
 
This new $N=2$ string  
was obtained by combining a conventional
$N=2$ string in one sector with a Green-Schwarz-Berkovits string in the 
other sector. The Berkovits string has an  $N=2$ superconformal invariance
but it is equivalent to the conventional $N=1$ $RNS$ string
through a non-local field redefinition 
(one first embeds the $N=1$ string into 
an $N=2$ string and then performs a non-local field redefinition that involves 
the $N=1$ ghost fields). So, in this sense our new "heterotic type $II$"
string is actually a new $(1,2)$ string. 
In the conventional heterotic $N=2$ strings one has to gauge 
a null current.
Depending on the null current one finds that the target space is effectively 
either two dimensional or three dimensional.  In our new $N=2$ string we 
also have to gauge a null current. 
We chose our null current such that it commutes 
with target space supersymmetry. With this choice the effective
target space in still four dimensional, but a fermionic 
direction is effectively gauged away, 
leading to self-dual superspace.

Let us indicate what happens if one had chosen a
fermionic spinor $u^{\alpha}$ instead of a bosonic one. 
A fermionic null spinor can always be written as $\lambda v^{\alpha}$,
where $\lambda$ is anticommuting and $v^{\alpha}$ is a bosonic
spinor. In order to obtain self-dual supergravity, the following 
conditions are sufficient: the vertex operators and gauge invariances
should not depend on $\lambda$, and $\lambda v^{\alpha}$ should be
covariantly constant.
It is not clear to us how the theory would implement the first
condition, and whether $\lambda$ is a new independent object or
should be expressed in terms of the world-sheet fields. 
Nevertheless, under these conditions all results in the
paper also apply to the case of a fermionic null spinor, except that
the central charge of the internal sector now has to be $12$ rather
than $6$. 

An interesting question is whether it is possible to formulate our model 
in $RNS$ variables.  One immediate problem is that space-time
fermions are very difficult to deal with in the $RNS$ formalism.
Conversely, it will be interesting to know  
whether one can translate the usual heterotic $(1,2)$ 
string with the conventional null vector in our formalism. 
Obviously, this is desirable only for the cases where the 
conventional $(1,2)$ strings do have target space supersymmetry.
One approach would be to follow the field redefinition from the 
$RNS$ string to the Green-Schwarz-Berkovits string. This involves, however,
non-local field redefinitions, so generically the null current 
will be non-local. It may be, though, that in the cases where 
the conventional $(1,2)$ strings do have target space supersymmetry
a yet new field redefinition exists such that the null current 
becomes local. This question definitely deserves further investigation.
 
In the usual $(1,2)$ strings
the four dimensional target space may be interpreted as a $(2,2)$-brane
moving in a 12 dimensional space-time of signature $(10,2)$ (thus,
indicating connections with $F$-theory). The new $N=2$ string we constructed
has as internal space an $N=2$ SCFT with central charge $c=6$ or $c=12$,
that typically involve $4$ or $8$ bosons.
It seems tempting 
to interpret our four dimensional target space as a $(2,2)$-brane
moving in a $(6,2)$ or $(10,2)$ space. Then   
the bosons would describe fluctuations in the transverse directions.
This interpretation becomes problematic, though, as soon as one computes the 
spectrum. Only half of those scalars survive the null projection. 
So, at least at first sight, it seems unlikely that such a picture
is correct\footnote{There still exists the possibility
that our model with $c=12$ describes  
an $N=1$ self-dual $(2,2)$-brane moving in the spacetime of 
12 dimensional $N=1$ self-dual supergravity in such
a way that the self-duality
effectively freezes half of the eight transverse directions. If such a
scenario is correct then the reduction from 12 to 4 dimensions  should 
be made in such a way that only an $N=1$ supersymmetry survives.
Similar ideas have been advocated by Ketov in \cite{k1}.
Let us also mention that our model may be related to 
theories in spacetimes with signature $(11, 3)$ proposed in \cite{seba} 
since we also have 3 timelike coordinates
(two non-compact and one compact, the $\r$ field).}.

An alternative way to derive the low energy effective action is 
to compute scattering amplitudes. For $N=2$ strings the most 
convenient way to do such calculations is to use the reformulation
as $N=4$ topological strings developed by Berkovits and
Vafa \cite{bv1} and subsequently used by Berkovits\cite{b1}, 
Ooguri and Vafa \cite{ov1}.
It will be interesting to perform
such a calculation not only in order to confirm the results we obtained
using sigma model methods but, more importantly, to study 
the internal sector, which is hard to analyze using  sigma 
model techniques.

One may use the techniques we developed in this article to construct 
new $N=1$ string theories. For instance, one could start from the 
type II string in the Green-Schwarz-Berkovits formalism\cite{bersie} (which is 
the $N=(1,1)$ model in the $RNS$ formalism), change the signature 
of space time to $(2, 2)$ and introduce a null current symmetrically 
in both sectors. A chiral null current contributes
$c=\pm 3$ to the central charge. Taking into account that the 
non-compact sector contributes another $c=-3$ and the $N=2$ ghosts $c=-6$,
we find that one would need an internal sector with $c=6$ or $c=12$. 
As such one could take a Calabi-Yau 2- or 4-fold.
Alternatively, one may gauge (symmetrically) a full $N=2$ null multiplet, 
in which case we would need a $c=15$ internal space which may be 
taken to be a Calabi-Yau 5-fold. Yet more models can be constructed
by asymmetric choices of the null currents. One can easily find appropriate
target space supersymmetric null currents for each of these cases.
All these models are difficult to construct in the $RNS$ formalism.
Altogether, we have seen many examples where the new Green-Schwarz-Berkovits
techniques are very powerful. In \cite{ogp} we witnessed string theory 
selecting a particular off-shell supergravity and in this article we
obtained a new $N=2$ string theory that has self-dual supergravity as
an effective field theory.   
It remains to be seen what the significance of these new $N=1$ models is.
We intend to return to these issues in the future. 

\section*{Acknowledgements}

We would like to thank N. Berkovits, 
D. Kutasov, E. Martinec, H. Ooguri, B. Peeters,
A. Van Proeyen, M. Ro\v{c}ek, 
C. Schweigert and  W. Troost 
for useful discussions. 
KS is supported by the European Commision HCM program CHBG-CT94-0734.
JdB is supported in part by NSF grant PHY-951497 and DOE grant
DOE-AC03-76SF00098, and is a fellow of the Miller Institute for Basic
Research in Science.

\appendix

\section{Conventions}

\renewcommand{\theequation}{A.\arabic{equation}}
\setcounter{equation}{0}

Let $y^1, y^2, y^3, y^4$ be real coordinates in a space of $(2, 2)$ signature,
\be
ds^2 = (dy^1)^2 + (dy^2)^2 - (dy^3)^2 - (dy^4)^2
\ee  
We define complex coordinates as follows
\bea
x^1 = \frac{1}{\sqrt{2}} (y^1 + i y^2); && 
x^2 = \frac{1}{\sqrt{2}} (y^3 + i y^4) \nonu
x^{\bar{1}} = \frac{1}{\sqrt{2}} (y^1 - i y^2); && 
x^{\bar{2}} = \frac{1}{\sqrt{2}} (y^3 - i y^4)
\eea
In these coordinates the line element is equal to 
\be
ds^2 = \h_{\ua \ub} dx^{\ua} dx^{\ub} 
\ee
where 
the underlined indices denote both an unbar and bar index, $\ua=(a, \ab)$,
(for spinors $\uua=(\a, \dda)$)  and
\be
\h_{\ua \ub} = \left(
\begin{array}{cc}
0 & \h_{a \bb} \\
\h_{\ab b} & 0 
\end{array}
\right)
\ee
with 
\be
\h_{a \bb} = \left(
\begin{array}{cc}
1 & 0 \\
0 & -1 
\end{array}
\right).
\ee  

The Dirac matrices in complex coordinates and for this specific 
complex structure have been worked out in \cite{kng1} (appendix B in the
second paper),
\be
\g^a = \left(
\begin{array}{cc}
0 & \s^a \\
\tilde{\s}^a & 0 \\
\end{array}
\right), \ \ 
\g^{\ab} = \left(
\begin{array}{cc}
0 & \s^{\ab} \\
\tilde{\s}^{\ab} & 0 \\
\end{array}
\right),
\ee
where the $\s$ matrices are given by the following matrices
($\s$ with upper indices and $\tilde{\s}$ with lower indices, see (\ref{vec}))
\bea
&& \s^1 = - i \sqrt{2} a(-,\dot{+}), 
\s^2 = - \sqrt{2} a(-,\dot{-}), 
\s^{\bar{1}} = - i \sqrt{2} a(+,\dot{-}), 
\s^{\bar{2}} = \sqrt{2} a(+,\dot{+}), \\
&& \tilde{\s}^1 = i \sqrt{2} a(\dot{-}, +),  
\tilde{\s}^2 = - \sqrt{2} a(\dot{+}, +),  
\tilde{\s}^{\bar{1}} = i \sqrt{2} a(\dot{+}, -),
\tilde{\s}^{\bar{2}} = \sqrt{2} a(\dot{-}, -), 
\eea
where the $a(\a,\b)$ denotes a $2 \times 2$ matrix 
with 1 at $(\a, \b)$ and zero elsewhere.
One may use these $\s$ matrices to convert vector indices to a pair
of two spinor indices. This is possible since  
$SO(2,2)=SL(2,R) \otimes SL(2,R)$.
Explicitly,
\be
V^{\a \dda} = {1 \over \sqrt{2}} V_{\ua} (\s^{\ua})^{\a \dda};
\ \ \ V^{\ua} = {1 \over \sqrt{2}} (\tilde{\s}^{\ua})_{\dda \a} V^{\a \dda}.
\label{vec}
\ee
In particular,
\be
V^{\a \dda} = \left(
\begin{array}{cc}
V_{\bar{2}} & -i V_{\bar{1}} \\
-i V_1 & - V_2 
\end{array}
\right) \label{transl}
\ee

One can check that the $\s$ matrices satisfy the following relations
\bea
&&{1 \over 2} \h_{\ua \ub} (\tilde{\s}^{\ua})_{\dda \a} 
(\s^{\ub})_{\b \ddb} = C_{\a \b} C_{\dda \ddb}, \\
&& {1 \over 2} (\s^{\ua})^{\a \dda} (\s^{\ub})_{\a \dda} = \h^{\ua \ub},
\eea
where $C_{\a \b}=C_{\dda \ddb}$ are  antisymmetric tensors
with $C_{+ -}=1$. We will raise and lower spinor indices with these
tensors. Our convention is the ``down-hill'' rule from the left to the 
right.  
These identities allow one to switch freely (with no extra factors) from 
one set of indices to the other. For example, one may check that 
\be
V^{\ua} U_{\ua} = V^{\a \dda} U_{\a \dda}
\ee  

{}From (\ref{transl}) we read off the $N=2$ complex conjugation rules for 
vectors
\be
\overline{V^{+\dot{+}}}=-V^{-\dot{-}}; \ \ 
\overline{V^{+\dot{-}}}=-V^{-\dot{+}}
\ee
In addition, we have
\be
\overline{D_{+}}=-D_{-}
\ee
and similar rules for the dotted indices. 

Our conventions for covariant derivatives, torsion, curvature, etc. 
are as follows
\begin{equation}
\del_A = E_A{}^{M} \pa_M + \ome_{A \b}{}^{\g} M_{\g}{}^{\b}
+ \ome_{A \ddb}{}^{\ddg} M_{\ddg}{}^{\ddb} + \G_A Y, \label{del} 
\end{equation}
\be
[ \del_A, \del_B \} = T_{AB}{}^{C}\del_C 
+ R_{AB\g}{}^{\d} M_{\d}{}^{\g} 
+ R_{AB\ddg}{}^{\ddd} M_{\ddd}{}^{\ddg} 
+ F_{AB} Y, \label{com}
\ee
where $E_A{}^{M}$, $\ome_{A \uub}{}^{\uug}$ and $\G_A$ are the vielbein,
the spin connection and the $U(1)$ connection, respectively. 
$T_{AB}{}^{C}$,  $R_{AB\uug}{}^{\uud}$ and $F_{AB}$ are the torsion,
the curvature tensor and the $U(1)$ curvature, respectively. 
$M_{\uud}^{\ \uug}$ are the 
generators of the Lorentz group.

\section{Analysis of VO's in components}

\renewcommand{\theequation}{B.\arabic{equation}}
\setcounter{equation}{0}

We show in this appendix, using a detailed component analysis, that the 
null current may be used to eliminate the chiral part of both the spacetime
and the internal sector vertex operators. We start from the latter.
By hitting equation (\ref{null}) 
with $D_{\dda}$ and going over to momentum space 
one gets
\be
u^\a p_{\a \dda} M_1 = 0
\ee
The general solution of this equation that does not set $M_1$ equal to zero
is 
\be
p_{\a \dda} = p' u_\a v_{\dda} \label{feq1}
\ee
where $p'$ is arbitrary and $v_{\dda}$ is a second commuting spinor. 
Clearly, this immediately implies that $p^2=0$.

The component expansion of $M_1$ is of the form 
\be
M_1 = a(x^-) + \q_\a \xi^\a (x^-) + \q^2 b(x^-)
\ee
where $x^-_{\a \dda} = x_{\a \dda} - i\q_\a \q_{\dda}$
(In our conventions $D_\a = \pa_\a - i \q^{\dda} \pa_{\a \dda}$, 
$D_{\dda} = \pa_{\dda} - i \q^{\a} \pa_{\a \dda}$).
Equation (\ref{internal}) implies the usual field equations 
for the component fields
\be
\Box a = \pa_{\a \dda} \xi^\a = b = 0 \label{field}
\ee
In addition (\ref{null}) yields 
\be
u^a \xi_\a =0,
\ee
which implies 
\be 
\xi^\a = \xi u^\a,  \label{feq2}
\ee
where $\xi$ is arbitrary anticommuting variable.
So the on-shell multiplet reads
\be
M_1 = a(x^-) + \q_\a u^\a \xi (x^-)
\ee
We will now show that the gauge invariance in (\ref{nuga}) is just enough 
to remove these degrees of freedom.
Indeed, starting from a general superfield $Y_1$ and imposing the 
condition that the $u^\a D_\a Y_1$ is chiral one obtains 
\be
\d M_1 = u^\a \psi_\a + \q^\a \q^{\dda} (- i \pa_{\a \dda} u^\b \psi_\b)
+ u^\a \q_\a T
\ee
where we have also used the on-shell condition (\ref{feq1}).
$\psi_\a$ is an arbitrary commuting spinor and $T$ is an arbitrary 
anticommuting variable. Clearly, these on-shell gauge transformations 
eliminate completely $M_1$. 
 
Let us now consider the 4D vertex operator. The component expansion 
of the gauge invariant field strengths are as follows 
\bea
&&W_\a = \l_\a + \q_\a D + \q^\b f_{\b \a} 
- i \q^2 \pa_{\a}{}^{\dda} \l_{\dda} \nonu
&&W_{\dda} = \l_{\dda} + \q_{\dda} D + \q^{\ddb} f_{\ddb \dda} 
- i \bar{\q}^2 \pa^{\a}{}_{\dda} \l_{\a}
\eea
where the same field $D$ appears in both expansions due to the 
identity $D^\a W_\a = D^{\dda} W_{\dda}$.
Equations (\ref{equa4D}) read
\bea
&&\pa_\a{}^{\dda} \l_{\dda} = \pa^\a{}_{\dda} \l_\a = 0, \\ 
&& \pa_{\a \dda} D = \pa_{\a \dda} f^{\a}{}_{\b} = 
\pa_{\a \dda} f^{\dda}{}_{\ddb} = 0
\eea 
Equation (\ref{null2}) implies (on-shell)
\be
u^\a \l_\a + u^\a \q_\a D + u^\a \q^\b f_{\a \b} = 0
\ee
This equation is then solved by 
\be
\l^\a = \l u^\a; \ \ D=0; \ \ f_{\a \b} = f u_{(\a} u_{\b)}
\ee 
where $\l$ and $f$ and on-shell fermionic and bosonic components.
So $W_\a$ on-shell reads
\be
W_\a = \l(x^-) u_\a + \q^\a f(x^-) u_{(\a} u_{\b)}.
\ee 

It is now easy to
show that the on-shell gauge invariance in (\ref{nuga1}) removes
these degrees of freedom. 
Indeed, working out (\ref{nuga1}) in components yields 
\be
\d W_\a = {1 \over 2} u_\a (F + \q^\b \pa_{\b \ddb} \chi^{\ddb} 
- i \q^\b \q^{\ddb} \pa_{\b \ddb} F)
\ee
which can clearly remove $\l$ and $f$ (in momentum space the $\q^\b$ term 
is proportional to $u_\a u_\b v_{\dda} \chi^{\dda}$). 

\section{Conservation of the supersymmetry currents.}

\renewcommand{\theequation}{C.\arabic{equation}}
\setcounter{equation}{0}

In this appendix we analyze the equations that follow by requiring 
conservation of the supersymmetry currents, $\pa H^{\pm}=0$.
The terms linear in the fermion field $\j^{\ua}$
have been analyzed in the main text.
The terms cubic in the fermions yield the following equations
\bea
&&(\h_{\ua \ue} \pm I_{\ua \ue}) (R_{\ue \ud \ub \uc} + 
2 H_{\uf [\ue \ub} H_{\uc] \ud \uf} 
+2 \del_{\ud} H_{\uc \ub \ue} - 2 \del_{\ue} H_{\uc \ub \ud}) - \nonu
&&\qquad 
- 6 H_{\ua \ud \ue} C^{\pm}_{\ue \ub \uc} + \del_{\ud} C^{\pm}_{\uc \ub \ua}
+ {\rm permutations \ \ with \ \ signs \ \ in} \ \ \ua, \ub, \uc =0
\label{Hcond1} \\
&&(\h_{\ua \ue} \pm I_{\ua \ue}) (R_{\ue \uud \ub \uc} +
2 \del_{\uud} H_{\uc \ub \ue}) + \nonu
&& \qquad +\del_{\uud}C^{\pm}_{\uc \ub \ua}
+ {\rm permutations \ \  with \ \ signs \ \ in} \ \ \ua, \ub, \uc=0,
\label{Hcond2} \\ 
&&(\h_{\ua \ue} \pm I_{\ua \ue})(\del_{\ue} T_{\uc \ub \uud}
-4 T_{\ub \uf \uud} H_{\uc \uf \ue}) -\nonu
&& \qquad -3 T_{\ua \ue \uud} C^{\pm}_{\ue \ub \uc}
+ {\rm permutations \ \ with \ \ signs \ \ in} \ \ \ua, \ub, \uc = 0.
\label{Hcond3}
\eea
We start our analysis with (\ref{Hcond2}). 
By using the identity
\be \label{cycl1}
R_{\uua[\ub\uc\ud]} = - 8 \nabla_{\uua} H_{\ub\uc\ud}
\ee
one finds that the complex structure independent part of 
$C^{\pm}_{\uc \ub \ua}$
is equal to $(-2/3) H_{\uc \ub \ua}$.
To obtain the complex structure dependent part of $C^{\pm}_{\uc \ub \ua}$
we observe that the term containing $I$ and the curvature $R$ in 
(\ref{Hcond2}) can be
rewritten as a linear combination of $I_{\ua\ue}R_{\uud[\ue\ub\uc]}$ and
$[\nabla_{\uud},\nabla_{\uc}]I_{\ua\ub}$. The first piece can be 
rewritten using the cyclic identity (\ref{cycl1}), the second piece using
(\ref{Jcond1}) and (\ref{Jcond2}). After some straightforward algebra one
obtains that the complex structure dependent part of $C^{\pm}_{\uc \ub \ua}$
is equal to $\pm I_{\ud [\uc} H_{\ub \ua] \ud}/3$.
Putting these results together we get that equation (\ref{Hcond2})
is solved by
\be
C^{\pm}_{\uc \ub \ua} = \frac{1}{3}(-2 H_{\uc \ub \ua} \pm 
I_{\ud [\uc} H_{\ub \ua] \ud}) \label{Cvalue}
\ee
 
An alternative route to obtain this result is to first use 
(\ref{Dcond1}) to eliminate the curvature from (\ref{Hcond2}) and 
then use the following identities involving torsions
\bea
&&T_{\uc \ub \dda} C_{\d \a} + {\rm cyclic \ \ in}\ \ \ua, \ub, \uc
= - i \del_{\d} H_{\ub \uc \ua} \\
&&C_{\a \d} T_{\ue \uc \dda} I_{\ue \ub} 
+ T_{\ua \uc \dde} C_{\d \e} I_{\ue \ub} - \ua \leftrightarrow \ub
= -i \del_{\d} (I_{\ud [\ua} H_{\ub] \uc \ud}) \label{Iident}
\eea
The latter is obtained by differentiating (\ref{Jcond2}) by $\del_{\d}$
and using 
\bea
T_{\d \dda, \uc, \dde} C_{\a \e} I_{\ue \ub} - \ua \leftrightarrow \ub
&=& (C_{\a \d} T_{\ue \uc \dda} I_{\ue \ub}
+T_{\ua \uc \dde} C_{\d \e} I_{\ue \ub}
+i \del_{\g} H_{\ddg(\d} I_{\a) \dda, \ub}
- \ua \leftrightarrow \ub) \nonu
&&-4i \del_{\g} H_{\ddg \d} I_{\ua \ub},
\eea 
which, in turn, can be obtained by using (\ref{id3}).

Equations (\ref{Hcond1}) and (\ref{Hcond3}) should now automatically 
be satisfied.
The complex structure independent part of (\ref{Hcond3}) reads
\be \label{aux01}
\del_{[\ua} T_{\ub \uc]}{}^{\uud} - 2 H_{\ue [\ua \ub} T_{\uc] \ue}{}^{\uud}
\ee
which is equal to zero by using the Bianchi identities.
The complex structure dependent part is equal to
\be
I_{[\ua|}{}^{\ue} \del_{\ue|} T_{\ub \uc] \uud} 
-2 I_{\ue \uf} H_{[\ub \uc}{}^{\ue} T_{\ua] \uf \uud}.
\ee
To show that this is equal to zero one first rewrites the derivative
of the torsion using (\ref{aux01}). Next one can ``partially
integrate'' the derivative in these terms using (\ref{Jcond3}),
after which one can use (\ref{Jcond2}) to work things
out and obtain zero. 

It remains to demonstrate (\ref{Hcond1}). This is most easily done
using the cyclic identity
\be \label{cycl2}
R_{\ua[\ub\uc\ud]} = - 4 \nabla_{\ua} H_{\ub\uc\ud}
\ee
which one can prove using the explicit form of $R$. 
It can be used to rewrite the complex
structure independent part of (\ref{Hcond1}) as a linear combination
of $\nabla_{[\ua}H_{\ub\uc\ud]}$ and $H_{\ue[\ua\ub}H_{\uc\ud]\ue}$.
The second of these terms vanishes identically, due to lack of
indices in dimension $4$, and the first term is then zero by virtue
of the Bianchi identity for $H$. The complex structure dependent
part of (\ref{Hcond1}) can be analyzed as follows. The term 
$I_{\ua\ue} R_{\ue\ud\ub\uc}$ can be rewritten as
a linear combination of  $I_{\ua\ue} R_{\ud[\ue\ub\uc]}$ 
and $[\nabla_{\ud},\nabla_{[\uc}]I_{\ub]\ua}$. The first
term can be manipulated using (\ref{cycl2}), the second using
(\ref{Jcond2}). After some manipulations we then end up
with a term proportional to $I_{\ua\ue} \nabla_{[\ue} H_{\ub\uc\ud]}$
and one proportional to $I_{\ua\uf} H_{\ue[\uf\ub} H_{\uc\ud]\ue}$,
and both vanish for the reasons mentioned above. 

\section{Beta-functions without the null current}

\renewcommand{\theequation}{D.\arabic{equation}}
\setcounter{equation}{0}

For completeness, we present here the full result of the beta-function
calculation. If we were to study the $(1,2)$-string with a different choice of
null current, one would first determine the additional constraints it implies
and subsequently insert them in the expressions below in order to 
determine the field equations for that string theory.

\bea
0 & = & R_{\uc\ub\ug\uf} H_{\uc\ua\ub} - T_{\uc\ub}{}^{\ua} T_{\ug\uf\ua} 
H_{\uc\ua\ub}
  + 2 \del_{\uc} T_{\ub\ug\uf} H_{\uc\ua\ub} \nonu
  & & + 4i T_{\uc\uf \a} T_{\uc\ug\dda} -2 T_{\uc\uf\ub} T_{\uc\ug\ud} 
H_{\ub\ua\ud}
  -2 T_{\uc\uf\ub} T_{\ud\ug\ub} T_{\ud\uc\ua} \nonu
  & & - 4i T_{\uc\ug \a} T_{\uc\uf\dda} +2 T_{\uc\ug\ub} T_{\uc\uf\ud} 
H_{\ub\ua\ud}
  +2 T_{\uc\ug\ub} T_{\ud\uf\ub} T_{\ud\uc\ua} \nonu
  & & -\frac{1}{4} T_{\ug\uf}{}^{\uub} R^{\uc}{}_{\uub\uc\ua}
   +\frac{i}{2} T_{\ug\uf}{}^{\a} T_{\a\ddb,\ua}{}^{\ddb} 
   +\frac{i}{2} T_{\ug\uf}{}^{\dda} T_{\dda\b,\ua}{}^{\b} \nonu
  & & -\frac{1}{2} T_{\uc\ua}{}^B R_{\uc B\uf\ug} 
  + \frac{1}{2} \del^{\uc} R_{\uc\ua\uf\ug} 
  -\frac{1}{2} R^{\uc}{}_{\ua\uc}{}^{\ub} T_{\ug\uf\ub} \nonu
  & & +\frac{1}{2} T_{\ue\ua\ub} T_{\ue\ub\uc} T_{\ug\uf\ua} 
  -\frac{1}{2} \del^{\uc} T_{\uc\ua}{}^{\ub} T_{\ug\uf\ub} 
  -T_{\uc\ua\ub} \del_{\uc} T_{\ug\uf\ub} \nonu
  & & + \frac{1}{2} \del^{\uc} \del_{\uc} T_{\ug\uf\ua} 
  - \frac{1}{2} T_{\ub\ua[\uf} T_{\ug]\ub}{}^{\uua} \del_{\uua}
    (\phi+\bar{\phi})
  + \frac{1}{2} \del_{\ua} T_{\ug\uf}{}^{\uua} \del_{\uua}
   (\phi+\bar{\phi})
\nonu
  & & +\frac{1}{2} T_{\ug\uf}{}^{\uua} \del_{\ua} \del_{\uua} (\phi+\bar{\phi})
  + \frac{1}{2} R_{\ud\ua\uf\ug} \del_{\ud} (\phi+\bar{\phi}) 
  + H_{\ue [ \ud\uf} H_{\ug] \ua\ue} \del_{\ud} (\phi+\bar{\phi}) \nonu
  & & + \del_{\ua} H_{\ug\uf\ud} \del_{\ud} (\phi+\bar{\phi}) 
  -\del_{\ud} H_{\ug\uf\ua} \del_{\ud} (\phi+\bar{\phi}) 
\nonu
0 & = & \frac{1}{2} R_{\uc\ub\ug\uf} T_{\uc\ub\uua} +
  \frac{1}{2} T_{\uc\ub}{}^{\ua} T_{\ug\uf\ua} T_{\uc\ub\uua} 
  - \del_{\uc} T_{\ub\ug\uf} T_{\uc\ub\uua} \nonu
  & & +  T_{\uc\ua[\uf} T_{\ug]\uc\ud} T_{\ua\ud\uua} 
  + 2 T_{\ua\uc[\uf} T_{\ug]\ud\uc} T_{\ud\ua\uua}
  +\frac{1}{2} \del^{\uc} \del_{\uc} T_{\ug\uf\uua} \nonu
  & & - \frac{1}{2}
    T_{\uf\ub\uua} T_{\ug\ub}{}^{\uud} \del_{\uud} (\phi+\bar{\phi})
  +\frac{1}{2} \del_{\ud} T_{\ug\uf\uua} \del_{\ud} (\phi+\bar{\phi})
  - T_{\uf\ue\uua} H_{\ug\ue\ud} \del_{\ud} (\phi+\bar{\phi}) \nonu 
  & & +\frac{1}{2}
      T_{\ug\ub\uua} T_{\uf\ub}{}^{\uud} \del_{\uud} (\phi+\bar{\phi})
  + T_{\ug\ue\uua} H_{\uf\ue\ud} \del_{\ud} (\phi+\bar{\phi}) 
\nonu
0 & = & \frac{1}{2} \del^{\uc} R_{\uc\uua\uf\ug} 
  -\frac{1}{2} R^{\uc}{}_{\uua\uc\ub} T_{\ug\uf\ub}
  +\frac{1}{2} \del_{\uua} 
  T_{\ug\uf}{}^{\td} \del_{\td} (\phi+\bar{\phi}) \nonu
  & & -\frac{1}{2} T_{\ug\uf}{}^{\uud}
  \del_{\uua} \del_{\uud} (\phi+\bar{\phi})
  + \frac{1}{2} R_{\ud\uua\uf\ug} \del_{\ud} (\phi+\bar{\phi})
  + \del_{\uua} H_{\ug\uf\ud} \del_{\ud} (\phi+\bar{\phi})
\nonu
0 & = & 2 \del_{\a} H_{\dda\b} - \del_{\b} \del_{\ua} (\phi+\bar{\phi})
\nonu
0 & = & 2 \del_{\dda} H_{\ddb\a} + \del_{\ddb} \del_{\ua} (\phi+\bar{\phi})
\eea

\end{document}